\documentclass[aps,pra, twocolumn, noshowpacs, floatfix]{revtex4}
\usepackage{}

\usepackage{amsfonts}
\usepackage{amssymb}
\usepackage{graphicx}
\usepackage{amsmath}
\usepackage[english]{babel}
\usepackage{color}
\usepackage{epstopdf}
\usepackage{footnote}

\begin{document}

\title{Optical-cavity manipulation strategies of singlet fission systems mediated by conical intersections: insights from fully quantum simulations}

\author{Kewei Sun$^{1,2}$, Maxim F. Gelin$^{1,2}$, Kaijun Shen$^{2}$, and Yang Zhao$^{2}$\footnote{Electronic address:~\url{YZhao@ntu.edu.sg}}}
\affiliation{$^{1}$\mbox{School of Science, Hangzhou Dianzi University, Hangzhou 310018, China}\\
$^{2}$\mbox{School of Materials Science and Engineering, Nanyang Technological University, Singapore 639798, Singapore} \\
}

\begin{abstract}

We offer a theoretical perspective on simulation and engineering of polaritonic conical-intersection-driven
singlet-fission (SF) materials. We begin by examining fundamental models, including Tavis-Cummings and Holstein-Tavis-Cummings Hamiltonians,
exploring how disorder, non-Hermitian effects, and finite temperature conditions impact their dynamics, setting
the stage for studying conical intersections and their crucial role in SF.
Using rubrene as an example and applying the numerically
accurate Davydov-Ansatz methodology, we derive dynamic and spectroscopic responses of the system
and demonstrate key mechanisms capable of SF manipulation, viz. cavity-induced enhancement/weakening/suppression of SF,
population localization on the singlet state via engineering of
the cavity-mode excitation, polaron/polariton decoupling, collective enhancement of SF.
We outline
unsolved problems and challenges in the field, and share our views on the development of the future
lines of research. We emphasize the significance of careful modeling of cascades of polaritonic conical
intersections in high excitation manifolds and envisage that collective geometric phase effects may
remarkably affect the SF dynamics and yield. We argue that microscopic interpretation of the main
regulatory mechanisms of polaritonic conical-intersection-driven SF can substantially deepen our
understanding of this process, thereby providing novel ideas and solutions for improving conversion
efficiency in photovoltaics.
\end{abstract}
\date{\today}
\maketitle

\section{INTRODUCTION}

Cavity molecular science is a rapidly developing interdisciplinary field that has gained significant attention in recent years\cite{Galego,Ribeiro2018,Mandal}. It explores the interaction between molecules and confined electromagnetic fields, such as those within optical or plasmonic cavities. By leveraging the strong coupling between light and matter and the highly hybridized electron-photon-phonon states thus formed, this field aims to manipulate chemical reactions, energy transfer processes, and molecular dynamics in novel manners. Applications have been found in diverse scenarios such as photo isomerization, singlet fission (SF), catalysis enhancement, selective bond manipulation, and intracavity electron transfer processes~\cite{Orgiu,Shalabney1,Riedinger,Bienfait,Vendrell}.
A typical example is the mixed light-matter quasi particles formed when the interaction between excitons and photons is strong enough, known as polaritons. In 1997, Agranovich {\it et al.~}proposed a theory of strong light-matter coupling in organic molecules~\cite{Agranovich}, and in 1998, Lidzey {\it et al.~}conducted the first experimental verification of this theory~\cite{LIDZEY1,LIDZEY2}.
In 2011, Schwartz et al.~\cite{Schwartz} introduced a photochemical coupling mechanism for optical switches under strong light, marking the beginning of advancements in polariton optics. Recent research has revealed that the strong coupling between molecular vibrations and a cavity mode can modify the rate constant of ground-state reactions, even in the absence of external driving. A notable characteristic observed in these experiments is the alteration of the vibrational spectrum, attributed to the formation of molecular vibrational polaritons -- a hybrid state arising from the interaction between molecular cavity modes and vibrational modes. The existence of vibrational polaritons and changes in chemical reactivity can be characterized by strong resonance~\cite{Kowalewski23}. Thus, controlling light-matter coupling can be leveraged for polariton transport engineering and the optimization of various photo-induced reactions~\cite{Vendrell,JYZ23a}.

\begin{figure*}[tbp]
\begin{minipage}[t]{0.4\linewidth}
\centering
\includegraphics[scale=0.168,trim=0 0 0 0]{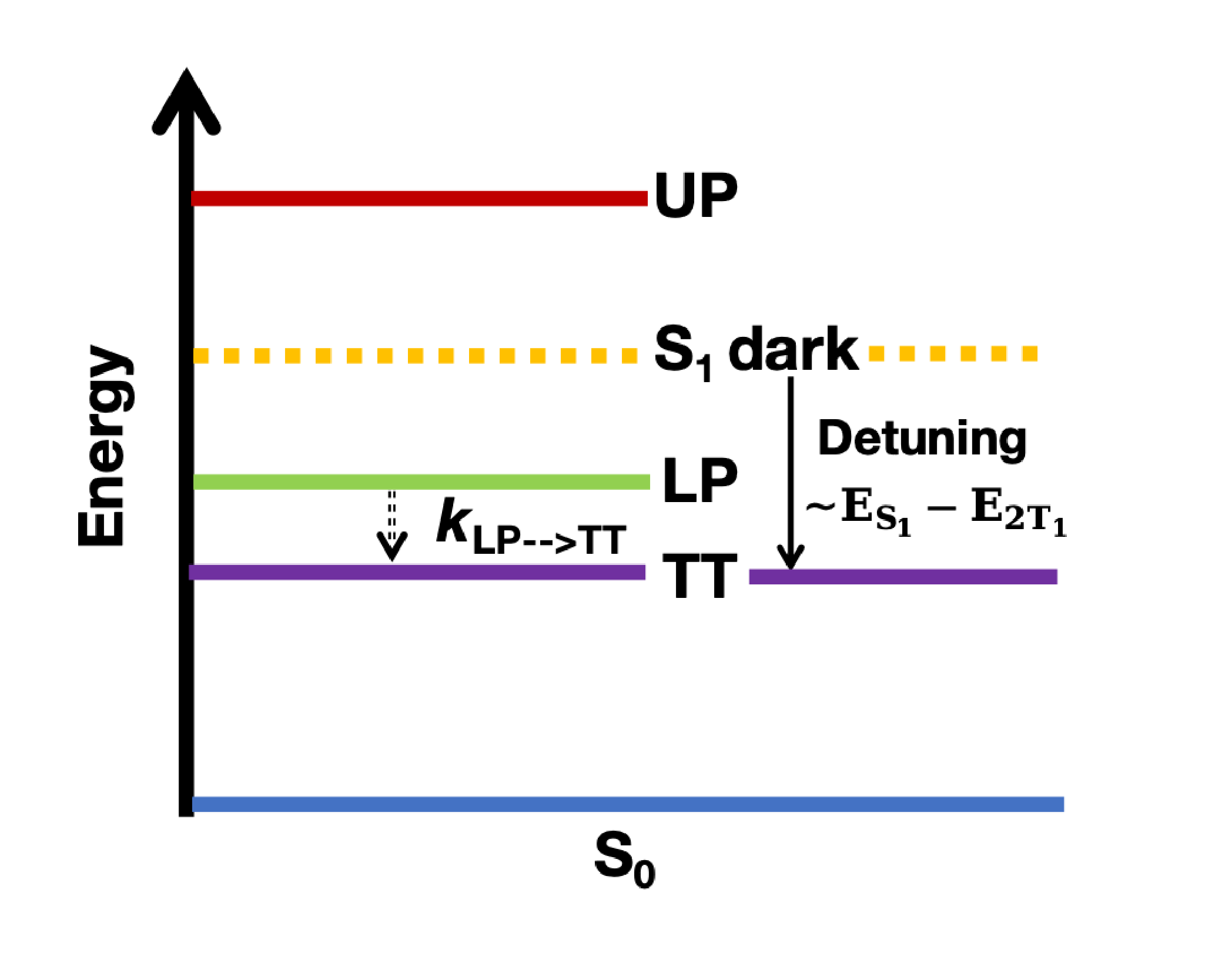}
\end{minipage}
\begin{minipage}[t]{0.4\linewidth}
\centering
\includegraphics[scale=0.2,trim=0 0 0 0]{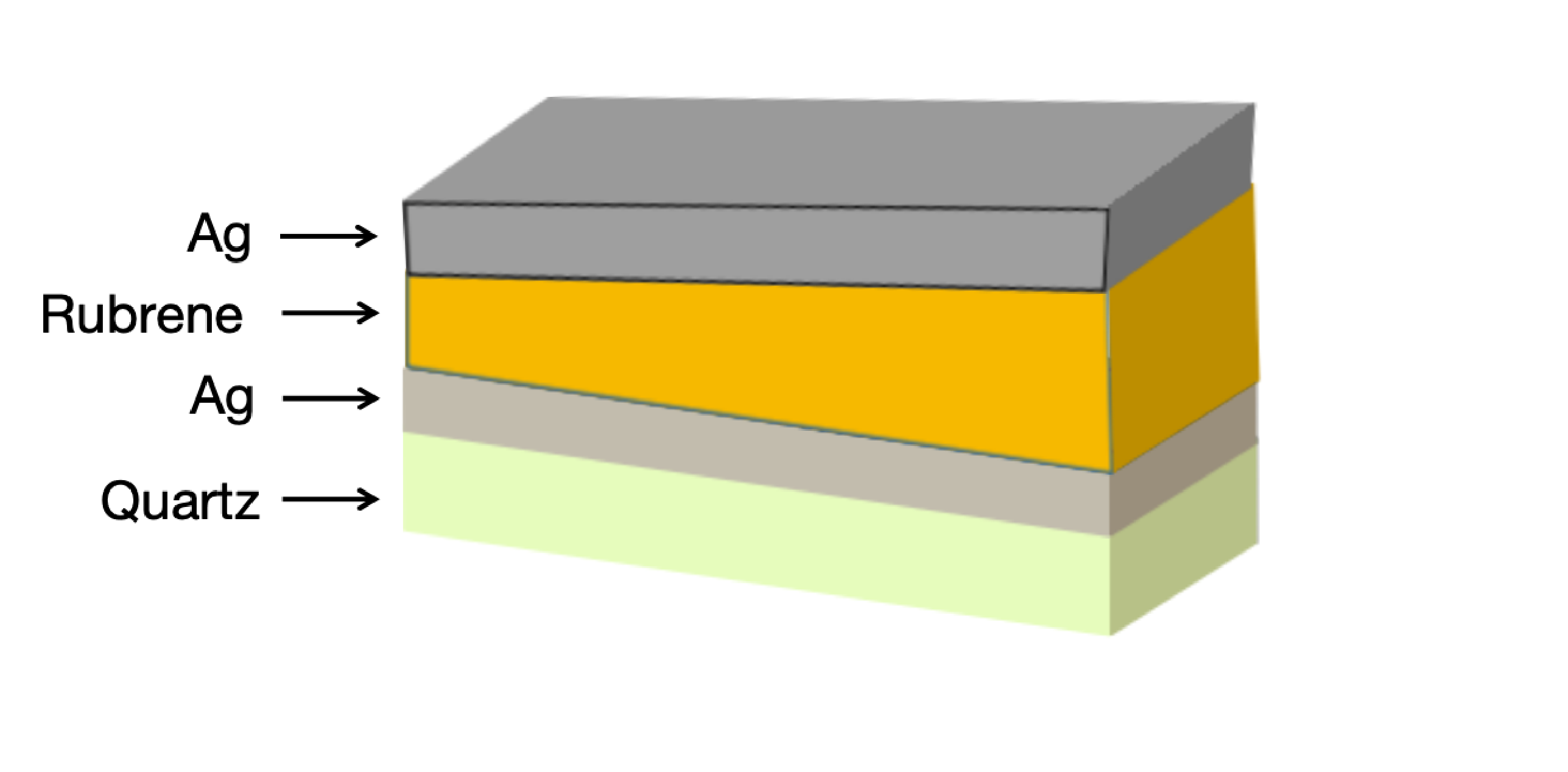}\\
\end{minipage}
\caption{PPP-mediated SF. A collective
bright singlet state coupled to the cavity mode forms the LP and UP states, while the collective singlet dark states are decoupled from the cavity
mode. (a) Energy diagram illustrating the initial step of the SF process in the presence of collective strong coupling. The LP efficiently channels population towards the TT state. (b) A schematic illustration of the slanted rubrene microcavity \cite{Takahashi}.}
\label{Fig0}
\end{figure*}

Six decades ago, Jaynes and Cummings drew inspiration from studies on quantum coherence in nuclear magnetic resonance and the solution to the Schr\"{o}dinger equation for two-level systems in a rotating magnetic field to explain deviations between Fermi's Golden Rule and perturbation theory in microwave resonators. This aligned with Rabi's findings in 1937. As a result, the renowned Jaynes-Cummings (JC) model~\cite{JAYNES} was developed, describing the interaction between a single two-level atom and a single-mode cavity field under the rotating-wave approximation (RWA). Despite its simplicity, the JC model offers an elegant framework for understanding light-matter interactions.
In recent years, various experimental setups have been devised to implement such a system~\cite{BIRNBAUM,MAUNZ,WALLQUIST}. Further research has revealed that the JC model exhibits significant complexity in terms of nonlinearity, leading to the development of various extended JC models. Among them, the most noteworthy are the Tavis Cummings (TC) model~\cite{TAVIS} and its variant without RWA, namely, the Dicke model~\cite{DICKE}. One of the most representative features of these Hamiltonians is a second-order quantum phase transition (QPT) from the normal phase to the super-radiant phase in the thermodynamic limit~\cite{BAUMANN,KLINDER,BADEN}, a collective phenomenon characterized by condensation of a macroscopic number of photons or atoms.

The interaction between the system and vibrational degrees of freedom (DOFs) can influence molecular behavior within the optical cavity, necessitating the incorporation of phonon modes into the TC model to create a composite framework. This composite model, known as the Holstein-Tavis-Cummings (HTC) model, is widely utilized to study various aspects of polaritonic transport. By tuning the interaction between atoms and cavity fields, steady-state exciton transport can be achieved, and factors influencing atomic transitions, as well as photon emission and absorption processes, can be examined. For the more complex HTC model, approximate analytical and numerical methods are typically employed to investigate its dynamic properties~\cite{LIU,Takahashi,WU,Spano15,Herrera2016,HERRERA2,HERRERA3,Cwik,HERRERA4}.

Strong light-matter coupling has been experimentally demonstrated for chromosomes of photosynthetic bacteria~\cite{Coles}, photosynthetic complexes~\cite{Tsargorodska}, and even for living Chlorobaculum tepidum bacteria~\cite{InVivo}. These and other phenomena demonstrate that polaritonic photophysics and photochemistry (PPP) holds potential for exciting practical applications, e.g., for engineering novel quantum materials~\cite{CavityMaterials} and development of polaritonic devices~\cite{KC16}. One of such applications, which holds a great promise for artificial light harvesting~\cite{Artificial22,Artificial22a} and can be substantially improved by the PPP machinery, is SF, a spin-allowed photophysical process discovered over fifty years ago in which a single photoexcited singlet exciton is converted into two triplet excitons (TT)~\cite{SF1,Smith2010,Smith2,Rev1,Rev2,JCP_SF,Rev3,Musser,HGD20,Miyata}.
By effectively doubling the number of generated excitons from one photon, SF offers a pathway to exceed the Shockley-Queisser limit and thus enhance power conversion efficiency in solar cells~\cite{Rao2017,Smith2010,Congreve2013}. As an all-organic and cost-effective approach to broaden spectral utilization, SF has stimulated extensive research aimed at understanding its underlying principles, optimizing its yields, and integrating it into diverse molecular architectures and solid-state platforms~\cite{Berkelbach2013,Margulies2016}. Nonetheless, controlling the rate and coherence of SF remains challenging, partly due to the intricate interplay of electronic interactions, vibronic couplings, and complex many-body dynamics~\cite{Monahan2015,Margulies2016}.

To transcend these molecular-level constraints, new strategies beyond chemical modification are being explored. One particularly promising route is to incorporate SF-active materials into optical cavities, drawing on concepts from polaritonic chemistry and cavity quantum electrodynamics~\cite{Mart,Schachenmayer,Ebbesen2016,Ribeiro2018,Flick2017,Hertzog2019}. Embedding SF systems inside cavities forms polaritons hybrid light-matter states that reshape exciton energy landscapes and relaxation channels~\cite{Munkhbat2018,Herrera2016,Galego2017}. Figure~\ref{Fig0} sketches the essentials of the cavity-enhanced SF, showing the lower polaritonic (LP) state and the upper polaritonic (UP) state producing via the coupling of the SF dimer to the cavity. Considering the large density of states in the manifolds of dark and TT states, it is essential that the TT states have to be energetically lower than the dark states, in order to prevent population leakage and enhance the SF yield (panel a). Hence the LP state acts as a new channel towards the TT state to accelerate SF. Figure~\ref{Fig0}(b) depicts the optical microcavity, in which the rubrene amorphous film is sandwiched between two silver mirrors for producing polaritonic states \cite{Takahashi}. By constructing the slanted rubrene microcavity, Takahashi {\it et al.} experimentally demonstrated the polaron decoupling effect \cite{Takahashi}  theoretically predicted by Spano \cite{Spano15}. They also found three critical parameters governing the cavity-mediated SF,  the excitonic fraction in the LP state, the energy alignment between the LP and TT states, and the Frank-Condon factor for the excitation~\cite{Takahashi}. Furthermore, through tuning cavity modes, detuning frequencies, and manipulating Rabi splittings, it becomes possible to influence electron-phonon couplings, reduce nonradiative losses, and reorient pathways critical to SF efficiency~\cite{Vendrell3,Tempelaar2017,Simpkins2021,CavitySF}.

Moreover, conical intersections (CIs) topological intersections of excited-state potential energy surfaces~\cite{b21,b22,b23,DS} play a crucial role in regulating SF by determining how excitons evolve and split~\cite{Berkelbach2013,Monahan2015,Musser,Miyata}. Under strong light-matter coupling, CIs can be effectively modified or navigated to favor SF channels~\cite{Gu2,sun1}. Such cavity-mediated modifications complement molecular tailoring: rather than relying solely on molecular packing or orbital engineering, one can now exploit photonic environments to dynamically adjust exciton dynamics, stabilize intermediate states, and potentially boost SF yields~\cite{Hertzog2019,Ribeiro2018}.

This Perspective focuses on how fully quantum simulations of dynamic observables and nonlinear spectroscopic signals elucidate the role of CIs and optical cavities in governing SF processes. By capturing the interplay of electrons, nuclei, and photons within a single quantum framework, these simulations provide new insights into cavity-mediated SF manipulation. They point toward cavity architectures that guide SF along desired pathways, offer greater resilience against disorder and thermal effects, and open unprecedented avenues for pushing SF efficiency beyond what is achievable in standard conditions~\cite{Vendrell3}. Ultimately, this integration of SF systems with cavities promises to inform both fundamental research and technological innovations, facilitating devices with improved energy conversion, increased stability, and enhanced coherence lifetimes.

The remainder of this Perspective is structured as follows.
Sec.~II introduces the fundamental models of light-matter coupling, namely the TC and the HTC Hamiltonian.
In Sec.~II A, the dynamics and absorption spectra of the disordered TC and HTC models are analyzed. Sec.~II B examines the impact of photon loss on HTC model dynamics, while Sec.~II C explores the effects of temperature on HTC model dynamics and absorption spectra using the thermo-field dynamics (TFD) approach.
Sec.~III offers a theoretical perspective on simulations, focusing on CI landscapes (Sec.~III A), SF dynamics (Sec.~III B), and two-dimensional spectra (Sec.~III C), providing new avenues for engineering polarized CI-driven SF materials. Sec.~IV highlights unresolved issues, challenges, and future directions in polaritonic system simulations. Finally, the conclusions and discussions are drawn in Sec.~V.

\section{Dynamics of light-matter coupling models}

To lay the groundwork for more realistic Hamiltonians that capture the essential physics of cavity-assisted SF processes, we start with basic models of light-matter interactions, namely the TC and the HTC Hamiltonians. By examining these fundamental models,
exploring how disorder, non-Hermitian effects, and finite temperature conditions impact their dynamics, we set the stage for studying more complex models focusing on CIs and their role in SF.
The Hamiltonian of the TC model can be written as
\begin{equation}
\hat{H}_{\mathrm{\rm TC}}=\omega_c \hat{a}^{\dagger} \hat{a}+\sum_{n=1}^N\left[\omega_n \hat{\sigma}_n^{+} \hat{\sigma}_n^{-}+g_n\left(\hat{a}^{\dagger} \hat{\sigma}_n^{-}+\hat{a} \hat{\sigma}_n^{+}\right)\right],
\label{TC}
\end{equation}
where $\omega_c$ is the frequency of the cavity mode, $\omega_n$ is the leap frequency of the $n$th qubit, and $g_n \equiv \omega_{\mathrm{R}, n} / \sqrt{N}$ is the coefficient of qubit-cavity coupling in the case of a single quantum bit $(n=N=1) $. It is simplified to the Rabi frequency $\omega_{\mathrm{R}}$ . $\hat{a}^{\dagger}(\hat{a})$
denotes the photon generation (annihilation) operator, $\hat{\sigma}_n^{+}=\left|1_n\right\rangle\left\langle 0_n\right|\left(\hat{\sigma}_n^{-}=\left|0_n\right\rangle\left\langle 1_n\right|\right)$ is the
raising (lowering) Pauli operator on the two-energy level atoms.
The Dicke model without the RWA is given by
\begin{equation}
\hat{H}_{\mathrm{\rm Dicke}}=\omega_c \hat{a}^{\dagger} \hat{a}+\sum_{n=1}^N\left[\omega_n \hat{\sigma}_n^{+} \hat{\sigma}_n^{-}+g_n\left(\hat{a}^{\dagger}+\hat{a}\right)\left(\hat{\sigma}_n^{-}+ \hat{\sigma}_n^{+}\right)\right].
\end{equation}
The TC model and the Dicke model are very common tools for studying the strong coupling between atomic systems and cavity modes.

Vibrational DOFs play a vital role in accurately describing organic molecular cavity QED systems as the coupling between electronic states and lattice vibrations (phonons) is essential for understanding decoherence, emission properties, and the overall behavior of these systems. After adding a bath of phonons to $\hat{H}_{\mathrm{TC}}$, we arrive at the Hamiltonian for the
HTC model
\begin{align}\label{HTC}
\hat{H}_{\mathrm{HTC}}=&\hat{H}_{\mathrm{TC}}+\sum_k \omega_k \hat{b}_k^{\dagger} \hat{b}_k\nonumber\\
&-\frac{\lambda}{\sqrt{N}} \sum_k \sum_{n=1}^N \omega_k \hat{\sigma}_n^{+} \hat{\sigma}_n^{-}\left(e^{-i k n} \hat{b}_k^{\dagger}+e^{i k n} \hat{b}_k\right) ,
\end{align}
where $\omega_k$ and $\hat{b}_k^{\dagger}$ are the frequency and the production (annihilation) operator of the $k$th mode of the phonon, respectively. $\lambda$ is the diagonal coupling strength of the qubit-phonon.

\subsection{Dynamics of disordered TC and HTC models}

The TC model is analytically solvable since the total number of excitations $\hat{N}_{\rm ex}=\hat{a}^{\dagger}\hat{a}+\sum_{n=1}^{N}\hat{\sigma}_n^{+}\hat{\sigma}_n^{-}$ is conserved~\cite{Huo,HTC}. Some methods, such as Green's function method~\cite{HTC,Gera} and complete basis expansion method~\cite{Huo,Wierzchucka}, can be applied to obtain the its eigenstates and eigenenergies.
In this subsection, we adopt the Green's function method to analytically get the eigenstates of the TC model, and its dynamics are subsequently studied, where the presence of disorder gave rise to a variety of complex behaviors. For the HTC model, we utilize the numerically accurate Davydov-Ansatz method, which is confirmed by comparing with the standard basis-set method~\cite{HTC}, to investigate the evolution of photon dynamics.

Below we consider the special case of $N_{\rm ex}=1$ because the manifold of single-photon excitation is the most relevant for applications. We note that our methodologies presented in this Perspective can be readily extended to the manifolds of multi-photon excitations. Thus, the state vectors span the closed ($N$+1)-dimensional Hilbert space with the basis set $\{|1_c,{\bf 0}_{\rm qu}\rangle, |0_c,1^{1}_{\rm qu}\rangle,\cdots,|0_c,1^{n}_{\rm qu}\rangle,\cdots,|0_c,1^{N}_{\rm qu}\rangle\}$. Here we assume that the cavity  mode is initially excited to the state  $|1_c,{\bf 0}_{\rm qu}\rangle$, while the qubits are in their ground states.
The Green's function $\mathcal{\hat{G}}(z)$ of the TC model in the matrix form can be written as
\begin{align}\label{Green1}
\mathcal{\hat{G}}(z)=\begin{bmatrix} {\rm \hat{G}^{-1}_{ph}}(z) & -\bf{g} \\ -{\bf g}^{\rm T} & {\bf\hat{G}}_{{\rm qu}}^{-1}(z) \end{bmatrix}^{-1},
\end{align}
where ${\rm\hat{G}_{ph}}(z)=(z-\omega_c)^{-1}$ and $[{\bf\hat{G}}_{{\rm qu}}(z)]_{m,n}=\delta_{m,n}(z-\omega_n)^{-1}$ are, respectively, free  photonic and qubit Green's functions. The vector ${\bf g}=(g_1,g_2,\cdots,g_N)$ denotes the coupling between the photon and the qubits. The photonic Green's function is explicitly  expressed  as
\begin{align}\label{Green2}
\mathcal{\hat{G}_{\rm ph}}(z)=\mathcal{\hat{G}}_{1_c,1_c}(z)=\frac{1}{{\rm \hat{G}^{-1}_{ph}}(z)-{\bf g}{\bf\hat{G}}_{{\rm qu}}(z){\bf g}^{\rm T}},
\end{align}
with the self-energy
\begin{align}\label{Green3}
{\bf g}{\bf\hat{G}}_{{\rm qu}}(z){\bf g}^{\rm T}=\sum_{n=1}^N\frac{g_n^2}{(z-\omega_n)}.
\end{align}
The photonic Green's function $\mathcal{\hat{G}}_{1_c,1_c}(z)$ is  directly related to a number of physical quantities, such as
the transmission coefficient of the waveguide,  population evolution of the photonic mode, and the emission or absorption spectrum.
The off-diagonal matrix elements of $\mathcal{\hat{G}}(z)$ are given by the expression
\begin{align}\label{GreenN}
\mathcal{\hat{G}}_{{\bf 1_{\rm qu}},1_c}^{\rm T}(z)=\mathcal{\hat{G}}_{1_c,{\bf1}_{\rm qu}}(z)=\frac{{\bf g}{\bf\hat{G}}_{{\rm qu}}(z)}{{\rm \hat{G}^{-1}_{ph}}(z)-{\bf g}{\bf\hat{G}}_{{\rm qu}}(z){\bf g}^{\rm T}},
\end{align}
which can be used for the description of correlations between the photon and the qubits. The reader is referred to Appendix~\ref{Green0} for additional details on the analytic solution of the TC model based on the Green's function method.

\begin{figure*}[t]
\centering
\includegraphics[scale=0.45,trim=-60 100 0 0]{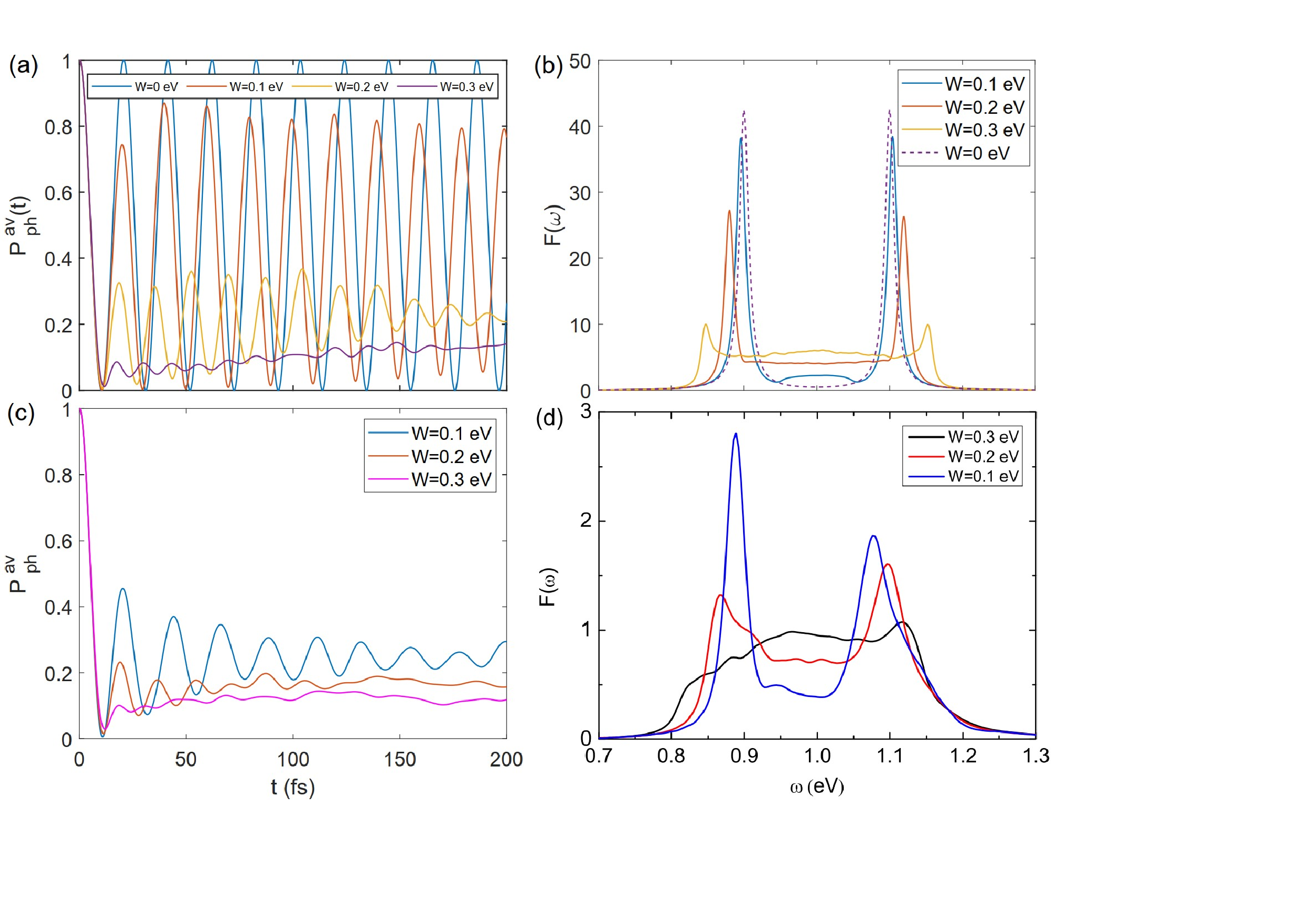}\\
\caption{(a) The photon-mode population dynamics evaluated for four diagonal disorder strengths indicated in the legend (the TC model with $N = 20$ qubits). (b) Absorption spectra of the TC model for three diagonal disorder strengths for a system $N = 100$ qubits. The other parameters are chosen as $\omega_c=1~\rm eV$, $\omega_{\rm R}=0.1~\rm eV$, $\omega_0=1~\rm eV$. 100 realizations of disorder are taken for obtaining converged results. (c) The photon-mode population dynamics of the HTC model with $\lambda= 0.5$ for three diagonal disorder strengths indicated in the legend. The central energy of the phonon band, $\omega_{k0} = 0.124~\rm eV$, is adopted. The remaining parameters are the same as in panel (a). (d) Absorption spectra calculated of the HTC model with $\lambda= 0.5$ for three values of energy disorder indicated in the legend. Adapted from \cite{HTC}. Copyright AIP Publishing.}
\label{Figure1}
\end{figure*}

Using Eq.~(\ref{Pop}), we study a disordered TC model with $N=20$ qubits. The photon-mode population dynamics for increasing disorder strengths ($W=0, 0.1, 0.2, 0.3 ~\rm eV$) are plotted in Fig.~\ref{Figure1}(a) without finite lifetimes.
The populations exhibit oscillatory behavior, and their median values decrease with static disorder. This can be understood as follows. Static disorder generates higher spread of the systems eigenenergies and richer energy spectrum. Hence more excitation remains localized in the qubits, because the back energy transfer to the photonic mode becomes hindered. Diagonal disorder thus promotes the energy transfer to the qubits. Furthermore, the  periods of  oscillations of the photonic-mode populations decrease with static disorder. The origin of this phenomenon is clarified by Eq.~(\ref{EG}) for the energy-gap frequency. Strictly speaking, Eq.~(\ref{EG}) is correct without disorder only. However, it remains qualitatively valid for non-too-high disorder if we average it over the relevant static-disorder distribution. Since $\omega_0$ enters the energy gap $\epsilon $ through $(\omega_0-\omega_c)^2$, this averaging can only increase the energy gap ($\epsilon_{\rm av } > \epsilon $) and therefore decrease the period of the energy-gap oscillations. Note also that  the short-time population dynamics is only slightly affected by diagonal disorder, because the difference between $\epsilon_{\rm av }$ and $\epsilon $ is relatively minor, and several periods are required to produce a substantial phase difference generated by the two frequencies.

To better understand the spectral features of the TC model, Fig.~\ref{Figure1}(b) shows absorption spectra for various disorder strengths for a cavity-QED system with 100 qubits. In the absence of disorder ($W=0$), the two peaks located at $\omega=0.9~\rm eV$ and $\omega=~1.1\rm eV$ are attributed to the bright eigenstates with the energies $E_{\pm}$, which are corresponding to the lower polariton (LP) and the UP peaks, respectively. The separation of the spectral peaks reveals the energy gap $\epsilon = 2\omega_{\rm R}=0.2$ eV, and the peak width is determined by the lifetime effect. As diagonal disorder removes degeneracy, the previously dark $N-1$ states gain certain oscillator strengths and form a new subband with the width $\sim W$ in the energy gap, as shown in Fig.~\ref{Figure1}(b).
{Similar conclusions were also reached in Refs.~\cite{Gerrit22a,Qiu2021,Houd}. {Specifically, for $g_n=g$ and site energies $\omega_n$ following a Gaussian distribution with a mean value $\omega_0$ and variance $\sigma$, the resonant gap between two polaritonic states, $2{\sigma^2}/(\sqrt{N}|g|)$, increases with
		disorder \cite{Gera1}.}
For an inhomogeneously broadened near-resonant system (with a weak disorder strength $W\ll\omega_0$), inhomogeneous broadening has no effect on the size of the splitting and, in general, does not lead to an inhomogeneous broadening of the split states. However, the width of the subband increases, the amplitudes of the two main peaks weaken, and inter-peak separation increases with static disorder.} These effects are the spectral counterparts of the disorder-induced decrease of oscillation periods and amplitudes of photon-mode populations discussed above. The contributions of the subband states to the absorbance are also enhanced due to the coupling to the cavity mode, leading to the emergence of the inter-band plateau whose value increases with disorder.
Diagonal disorder often leads to the broadening of absorption peaks. However, this effect is not pronounced in the present case: Increase of the subband plateau is accompanied by the overall spectral intensity weakening and shrinking the widths of the two main peaks.

Proper inclusion of vibrational modes is essential for a comprehensive and realistic depiction of organic molecular cavity QED systems.
By incorporating phonon modes and qubit-phonon couplings into Hamiltonian (\ref{TC}), we obtain Hamiltonian (\ref{HTC}) of the HTC model. Since the HTC Hamiltonian commutes again with the excitation number operator $\hat{N}_{\rm ex}=\hat{a}^{\dagger}\hat{a}+\sum_{n=1}^{N}\hat{\sigma}_n^{+}\hat{\sigma}_n^{-}$, we can work in the lowest-energy  excitation manifold and set $N_{\rm ex}=1$. To handle the newly introduced vibrational DOFs and to accurately solve the multidimensional time-dependent Schr\"odinger equation governed by the HTC Hamiltonian, we adopt the time-dependent variational method with a trial state from the hierarchy of Davydov Ans\"{a}tze ~\cite{25,26,27,28,29,yaoyao,30,skw20a}.
The multiple Davydov $\mathrm{D}_2$ Ansatz (also known as multi-$\rm D_2$ Ansatz) with multiplicity $M$ can be written as
\begin{equation}\label{D2}
|{\rm D}_{2}^{M}(t)\rangle=\sum_{n}|n\rangle\sum_{m=1}^{M}A_{mn}(t)e^{\left(\sum_{k}f_{mk}(t)\hat{b}_k^{\dagger}-\mathrm{H.c.}\right)}|0\rangle_{\mathrm{ph}},
\end{equation}
{where $|n\rangle$ stands for the ($N$+1)-dimensional basis state}, $|0\rangle_{\mathrm{ph}}$ is the phonon vacuum state, $\rm H.c.$ represents the Hermitian conjugate, $A_{mn}(t)$ are the amplitudes of the excited states, and $f_{mk}(t)$ are the harmonic mode displacements for the $m$th coherent state and the $k$th mode. {The multi-D$2$ Ansatz includes a superposition of multiple coherent states for each boson mode, which helps deliver accurate wave packet motion on the potential energy surfaces, thus leading to the convergent numerically accurate results for a sufficiently large multiplicity $M$. The approach employing the multi-D$_2$ Ansatz belongs to a large family of methods which use time-dependent Gaussian basis functions to accurately solve multidimensional time-dependent Schr\"{o}dinger equations.
The reader is referred to Appendix \ref{md2} for derivation details on the time evolution of the multi-D$_2$ Ansatz for the HTC model as well as calculations of related physical observables.

The time-dependent photon-mode populations of the HTC model are plotted in Fig.~\ref{Figure1}(c) for different diagonal disorder strengths indicated in the legend with the qubit-phonon coupling coefficient $\lambda=0.5$. A linear phonon dispersion is assumed to be of the form
\begin{equation}
\omega_k=\omega_{k0}[1+\Omega(\frac{2|k|}{\pi}-1)],
\end{equation}
where $\omega_{k0}$ denotes the central energy of the phonon band, $\Omega\in[0,1]$ is the band width, and the momentum is set to be $k=2\pi{l}/N$ with $(l=-\frac{N}{2}+1,\cdots,\frac{N}{2})$. Here the band width of the phonon spectrum is set to $\Omega=0.5$. The populations do not show any signatures of coherent vibrational effects and exhibit oscillations caused by the coupling of the photonic and qubit modes. The qubit-phonon coupling significantly reduces amplitudes of these oscillations. Furthermore, the photon-mode populations gradually approach steady-state values that decrease with $W$, indicating that stronger qubit-phonon coupling facilitates photon-to-phonon mode population transfer.
It is apparent that the qubit-phonon coupling plays a role similar to that of the diagonal disorder, in agreement with Ref.~\cite{GB21} which shows that static disorder can be handled on equal footing with the dynamic disorder caused by vibrational modes. From figs.~\ref{Figure1}(c), it reveals that strong diagonal disorder lead to the localization of qubit excitations.

The influence of the energy disorder on the absorption spectra for the HTC model with $\lambda=0.5$ is elucidated by Fig.~\ref{Figure1}(d). Qualitatively, disorder makes dark vibronic polariton states partially bright, owing to incomplete destructive interference. This produces new spectral features in the absorption spectra which are shown in Fig.~\ref{Figure1}(d) for various diagonal disorder strengths. The spectral plateau sandwiched between the two peaks gradually rises, and the polariton splitting widens with $W$. This is attributed to the disorder-induced dipole moment of the dark vibrational polariton. For $W=0.1~\rm eV$, the mirror symmetry with respect to $\omega =1$ eV is broken, i.e., so-called the polaron-decoupling effect. {The electronic and vibrational degrees of freedom contributing to the LP state become separable, which explains relative insensitivity of this peak to the qubit-phonon coupling. On the contrary, the UP peak is significantly broadened by increasing $\lambda$ and exhibits new spectral features. The LP peak largely retains its shape, while the amplitude, position, and width of the UP peak change markedly. Takahashi {\it et al.}~utilized the HTC model for the interpretation of their steady-state and time-resolved spectroscopic measurements. They demonstrated that when the cavity photon energy is altered, the rate of SF  resonantly excited into the LP state is indeed modulated. This finding implies that polaron decoupling is crucial in nonadiabatic dynamics~\cite{Takahashi}. In addition, the extent of polaron decoupling is highly sensitive to the ratio of Rabi splitting to the reorganization energy. Once this ratio surpasses a critical value, complete decoupling occurs, which in turn may modify the chemical reaction rates~\cite{Zeb18}}.

\subsection{Dynamics of the non-Hermitian HTC model}

Hamiltonian~(\ref{HTC}) is Hermitian, but in a realistic cavity system photon loss is inevitable due to the following factors: geometric deflection, diffraction, incomplete reflection of cavity mirrors, and non-activated absorption and scattering within materials, which greatly affects the behaviors of cavity assisted QED. To take into account
various dissipative effects that are not included in the reservoir
Hamiltonian $\hat{H}_{\rm R}$, such as cavity loss, a non-Hermitian term~\cite{zhang24},
\begin{equation}
\hat{H}_{\rm nH} = -i\kappa|1_c,{\bf 0}_{\rm qu}\rangle\langle 1_c,{\bf 0}_{\rm qu}|,
\end{equation}
can be added to Eq.~(\ref{HTC}), where $\kappa$ is the loss rate. The effective
Hamiltonian with the non-Hermitian term neglects, in comparison with the Lindblad master equation,
the fluctuation term $2\kappa \hat{a}\rho \hat{a}^{\dagger}$.
A more sophisticated scheme includes the coupling between
the cavity mode and the environmental modes (namely, the
Gardiner-Collett interaction Hamiltonian), characterized by a continuum
spectral density~\cite{Huo}. This Hamiltonian can be rigorously derived
from the QED first-principles. It has been used to investigate
polariton quantum dynamics in a dissipative cavity. By constructing
an effective non-Hermitian Hamiltonian or employing spectral
discretization techniques, the time-dependent variation method
with the mD2 Ansatz can be applied to account for dissipation
scenarios.

For a nonzero cavity loss rate, both the upper and lower polariton
states experience a depletion of their population, transferring it
to the ground state. This loss of the excited state population typically
diminishes the capacity of the system to undergo reactions on
excited surfaces. {However, cavity loss may also enhance the rate of photochemical
reactions~\cite{Torres}. Thus, cavity loss can be viewed as an additional adjustable element for controlling photochemical polaritonic reactivity.}

\begin{figure}[t]
\centering
\includegraphics[scale=0.4,trim=30 100 0 0]{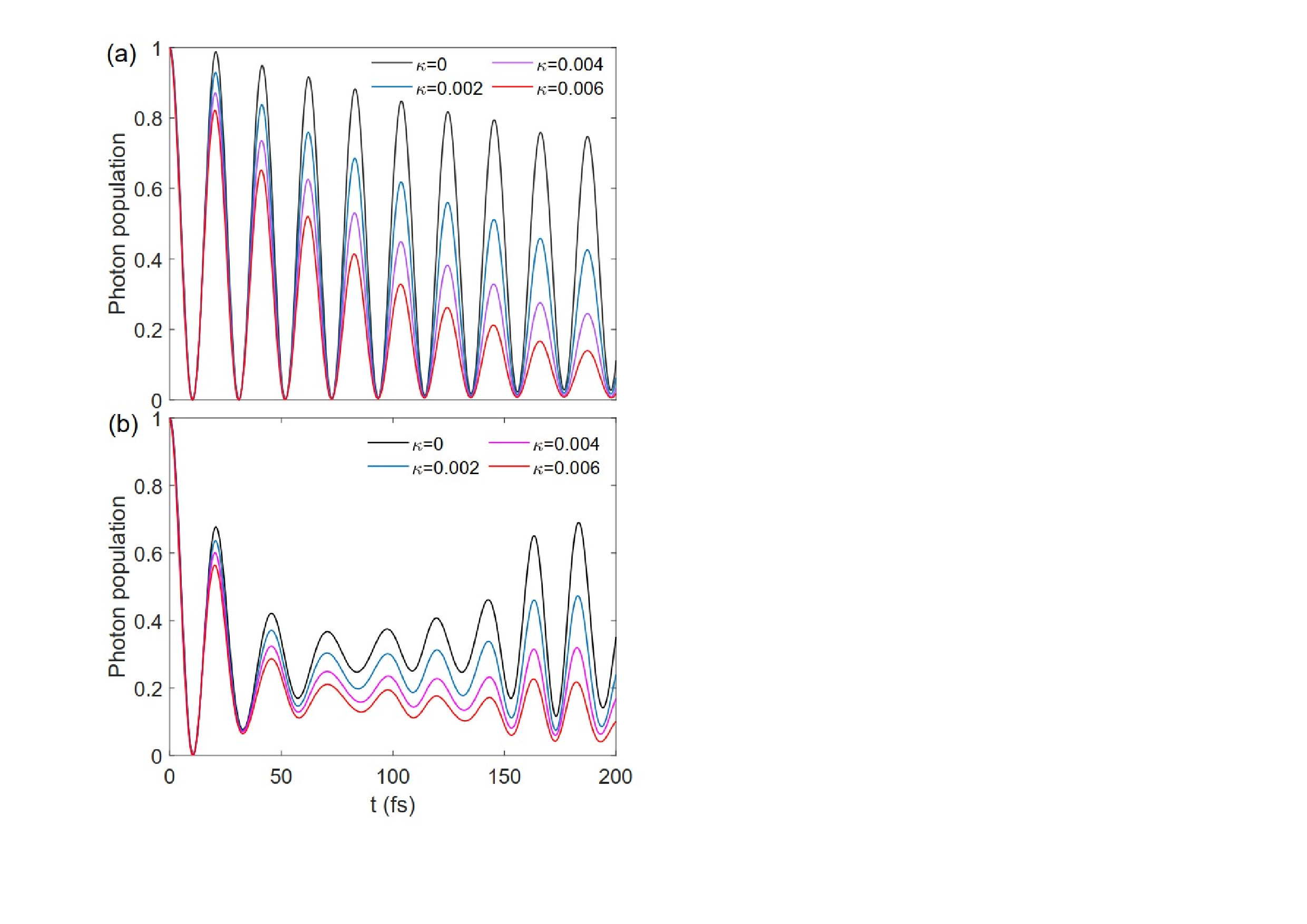}\\
\caption{The photon-mode population dynamics with the finite lifetimes $\kappa=0\rm eV, 0.002\rm eV, 0.004\rm eV,$ and 0.006 eV for $\lambda = 0.1$ (a) and $\lambda = 0.4$ (b). The other parameters are set by $\omega_R=0.1\rm eV$, $N=10$, $\omega_c=\omega_n=1\rm eV$, and $\Omega=0.5$. Adapted from \cite{zhang24}. Copyright AIP Publishing.}
\label{fig2}
\end{figure}

The photon population evolutions $\langle D_2^M(t)|\hat{a}^{\dagger}\hat{a}|D_2^M(t)\rangle$ for different
cavity loss parameters $\kappa$ and qubit-phonon couplings $\lambda$ are
plotted in Fig.~\ref{fig2}. As the cavity loss increases, the photon population
gradually decreases. Due to the finite lifetimes ($\kappa\neq0$), the population
decays to zero in the long-time limit. The bright-state energy
gap for the TC model is $\epsilon=\sqrt{(\omega_c-\omega_0)^2+4\omega_R^2}=2\omega_R$, which determines
the oscillation period $2\pi/2\omega_R\sim20.7~\rm fs$ of the photon population.
As a comparison, for the case of weaker qubit-phonon coupling in the HTC model as shown in Fig.~\ref{fig2}(a), we can still
observe the regular 20.7 fs periodic oscillations. For $\lambda= 0.4$, the
phonon-qubit coupling significantly reduces the amplitudes of these
oscillations, as shown in Fig.~\ref{fig2}(b). Stronger qubit-phonon coupling
quenches Rabi oscillations, due to the increasing number of
quantum states participating in the HTC dynamics. It is worth mentioning
that a partial recurrence of photon population occurs after
150 fs. We attribute it to the limited number of discrete phonon
modes in the model. In addition, the period of Rabi oscillations has
a small increment for a stronger phonon-qubit coupling strength in
Fig.~\ref{fig2}(b). Hence, the renormalized Rabi frequency is slightly reduced
by the qubit-phonon coupling.

\subsection{Finite-temperature dynamics of the HTC model}\label{HouE}

\begin{figure*}[tbp]
\includegraphics[scale=0.42,trim=20 220 0 0]{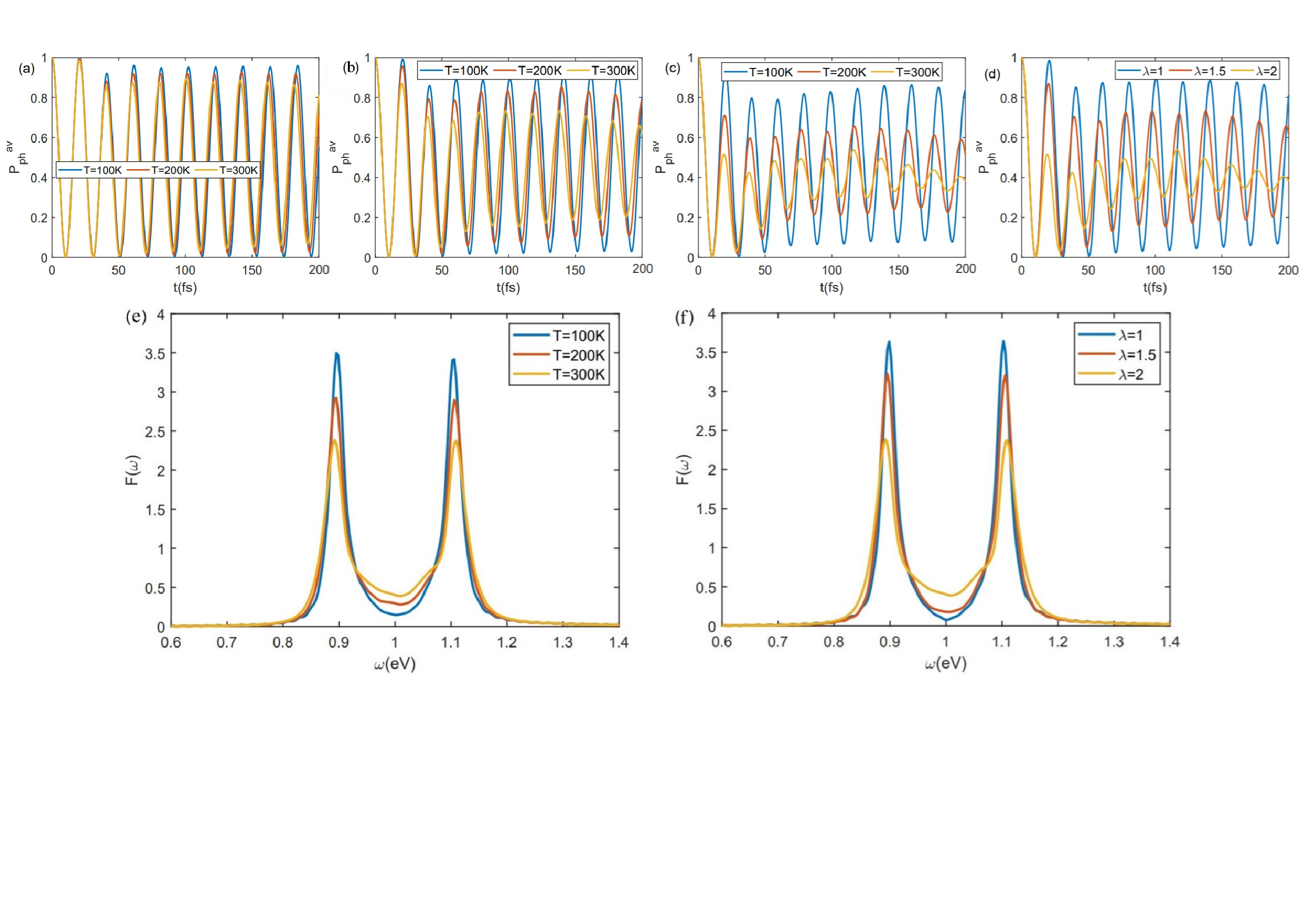}\\
\caption{Evolution of photon population for different qubit-phonon diagonal coupling strengths and temperatures. $\omega_{k0}=0.0124\mathrm{eV}$, $\omega_{\mathrm{c}}=1\mathrm{eV}$, $\omega_n=1\mathrm{eV}$
(a)$\lambda=1$, $T=100$K, $T=200$K, $T=300$K;(b)$\lambda=1.5$, $T=100$K, $T=200$K,$T=300$K;
(c)$\lambda=2$, $T=100$K, $T=200$K, $T=300$K;(d)$T=300K$, $\lambda=1$, $\lambda=1.5$, $\lambda=2$. Absorption spectra at different temperatures and qubit-phonon coupling strengths $\omega_{k0}=0.0124 \mathrm{eV}$. (e)$\lambda=2$, $T=100~\rm K$, $T=200~\rm K$, $T=300~\rm K$. (f)$T=300~\rm K$, $\lambda=1$, $\lambda=1.5$, $\lambda=2$.  Adapted from \cite{HouE24}. Copyright AIP Publishing.}
\label{figure 2}
\end{figure*}
Usually, ambient temperatures have a significant and multifaceted effect on polaritonic
transport, notably if it is driven by vibrational modes whose energies
are comparable with thermal energies. We can tackle this challenge by employing the TFD representation, which permits us to
treat quantum systems at finite temperatures by employing the TFD Schr$\ddot{\rm o}$dinger equations. The reader is referred to Appendix \ref{tfd} for the theoretical framework of the TFD representation in the time-dependent variational calculation with the multi-D$_2$ Ansatz.

In this subsection, we discuss the finite-temperature dynamics of the HTC model, Hamiltonian (3), in which only the diagonal qubit-phonon coupling is considered. Time evolutions of the photon populations for different coupling strengths and temperatures is show in Fig.~\ref{figure 2}. The temperature effect on the photonic dynamics is relatively weaker for smaller $\lambda$, as shown in Figs.~\ref{figure 2}(a)-(c). Coherent oscillation of the photon populations is gradually weakened with the increase of the temperature. Here the temperature effect is related to the qubit-phonon coupling. It is much clear if we roughly express the dressed qubit-phonon coupling strength as $\sum_{k}\lambda \cosh \theta_k+\sum_{k}\lambda \sinh \theta_k$, which increases with temperature, from 1 at $\rm T = 0$ to infinity as $\rm T\to\infty$. Obviously, the qubit-phonon coupling $\lambda$ plays the role of an amplification parameter. Besides, the expression of dressed qubit-phonon coupling also tells us that the qubit-phonon coupling acts similarly to temperature, which is further verified by Figs.~\ref{figure 2}(c) and (d). Hence, significance of temperature effects in the HTC model is determined by two conditions: (1) $k_BT$ is comparable with the phonon frequencies $\omega_k$. (2) $\lambda$ is not very weak~\cite{HouE24}.

The HTC absorption spectra without detuning are plotted in Fig.~\ref{figure 2}. There are two peaks, i.e., the LP and UP peaks, which are approximately located at $0.9\rm eV$ and $1.1\rm eV$, respectively. They are mainly attributed to the bright states with energies $E_{\pm}=1/2(\omega_c+\omega_0)\pm1/2\sqrt{(\omega_c-\omega_0)^2+4\omega_R^2)}$. The separation of the spectral peaks reveals the energy gap $\epsilon\sim2\omega_R = 0.2~\rm eV$. As the temperature increases, the peak intensities are weaken, and the separation of the spectral peaks slightly increase due to enhanced dynamics disorder, as shown in Fig.~\ref{figure 2}(e). It is worth mentioning that the middle portion of the absorption spectrum between the peaks gradually increases in height with the increasing temperature due to the peak broadening. As expected, similar spectral behaviors are exhibited in Fig.~\ref{figure 2}(f) for different qubit-phonon coupling strengths. Since smaller Huang-Rhys factors are taken, symmetric two peaks are found even though the polaron-decoupling threshold $2N\omega_R/\lambda^2\omega_k=1$ of the LP state is satisfied.

{Our results are consistent with the observation by Laitz {\it et al.}~that exciton-phonon scattering decreases as the temperature drops. This decrease leads to an alteration of the bottleneck effect in cavity-controlled perovskite films. Simultaneously, by adjusting the detuning of the microcavity, the optical properties of the perovskite films, such as photoluminescence  quantum yield, are greatly changed. Obviously, the temperature control and the cavity detuning can be regarded as the new ``knobs" to regulate the photochemical process~\cite{Laitz2023}. However, the open question is how temperature-dependent polariton relaxation mechanisms  affect the rate and product selectivity of polariton-mediated chemical reactions.}

\section{CI-mediated SF in optical cavities}

Our aim in this section is to highlight the status quo of the studies into the CI-driven cavity-mediated SF from the point of view of theoreticians.
To enhance the readability of this work, a bottom-up approach is adopted. Specifically, we focus on a single representative SF species, rubrene dimers (RDs), and conduct an in-depth analysis of a well-characterized QED system (two RDs in a single-mode cavity).
Furthermore, we explicitly monitor the first (ultrafast) leg of the SF process, in which the singlet exciton produces  a correlated triplet exciton pair. The process that follows transforming the correlated triplet pair into free triplet states occurs on a much longer timescale and is not considered.
We start by reviewing the fundamental processes that shape SF mechanisms and highlighting their spectroscopic signatures. Subsequently, we place these findings within the broader context of theoretical investigations into PPP molecular systems and materials, outlining open questions and computational challenges on the horizon.
All simulations here have been performed using time-dependent variation with the multi-D$_2$ Ansatz (cf.~Appendix C), a well-established many-body wave-function-based variational integrator of quantum dynamics employing Gaussian coherent states \cite{zhao2,zhao1}. Provided that the multiplicity of the multi-D$_2$ Ansatz is sufficiently high (this is so for all numerical illustrations of the present work), the trial state can be considered numerically accurate, effectively capturing and faithfully reproducing the dynamics of many-body quantum systems with multiple DOFs.

A molecular (hence the subscript M) system consisting
of two rubrene dimers coupled to a cavity photonic mode described
by the Hamiltonian
\begin{equation}
H=\sum_{j=1,2}H_{{\rm M}}^{(j)}+H_{{\rm C}}+H_{{\rm CM}}^{(j)}.\label{Htot}
\end{equation}
Here $H_{{\rm M}}^{(j)}$ refers to the $j$th pristine (without cavity)
dimer, $H_{{\rm C}}$ specifies the cavity field, and $H_{{\rm CM}}^{(j)}$
is responsible for the dimer-cavity photonic coupling. In the diabatic
representation, each dimer is modeled as a five-electronic-state two-vibrational-mode
system~\cite{sun2}:
\begin{eqnarray}\label{HM}
H_{{\rm M}}^{(j)}& = & \sum_{k={\rm g}^{(j)},{\rm S}_{1}^{(j)},{\rm TT}^{(j)},{\rm S}_{n}^{(j)},{\rm TT}_{n}^{(j)}}|k^{(j)}\rangle(\varepsilon_{k}+h_{k}^{(j)})\langle k^{(j)}| \nonumber \\
& + & (|{\rm S}_{1}^{(j)}\rangle\langle{\rm TT}^{(j)}|+|{\rm TT}^{(j)}\rangle\langle {\rm S}_{1}^{(j)}|)\lambda Q_{cu}^{(j)}.
\end{eqnarray}
Here $|{\rm g}^{(j)}\rangle$ denotes the electronic ground state, $|{\rm S}_{1}^{(j)}\rangle$
is the singlet electronic state coupled to the triplet-pair state
$|{\rm TT}^{(j)}\rangle$, while $|{\rm S}_{n}^{(j)}\rangle$ and $|{\rm TT}_{n}^{(j)}\rangle$
are higher-lying singlet and triplet states. $\lambda$ is the coupling
constant specifying the $|{\rm S}_{1}\rangle$-$|{\rm TT}\rangle$ CI responsible
for the SF, $\varepsilon_{k}$ are the electronic vertical excitation
energies, and $h_{k}^{(j)}$ are the vibrational Hamiltonians involving
a single interstate coupling mode (subscript $cu$) and a single primary tuning
mode (subscript $tu$):
\[
h_{k}^{(j)}=h_{g}^{(j)}+\kappa_{k}Q_{tu}^{(j)},
\]
\[
h_{g}=\frac{1}{2}\sum_{\alpha=cu,tu}\omega_{\alpha}([P_{\alpha}^{(j)}]^{2}+[Q_{\alpha}^{(j)}]^{2})
\]
($\hbar=1$). Here $Q_{\alpha}^{(j)}$ and $P_{\alpha}^{(j)}$ are
dimensionless coordinates and momenta of the interstate coupling ($\alpha=cu$)
and tuning ($\alpha=tu$) modes with frequencies $\omega_{\alpha}$,
and $\kappa_{k}$ are the linear intrastate electron-vibrational coupling
constants. The effective model of the CI-mediated SF process has been
developed in Ref. \cite{Miyata} and further extended in Refs. \cite{Sun6,skw}.
In this model, the tuning mode $Q_{tu}$ is the dimensionless reaction
coordinate of the SF process, while $Q_{cu}$ is an antisymmetric
interstate coupling mode which couples the states $|{\rm S}_{1}^{(j)}\rangle$
and $|{\rm TT}^{(j)}\rangle$. This model is augmented
with the higher-lying excited electronic states $|{\rm TT}_{n}^{(j)}\rangle$ and
$|{\rm S}_{n}^{(j)}\rangle$.

The cavity-mode Hamiltonian reads
\[
H_{{\rm C}}=\omega_{{\rm C}}C^{\dagger}C
\]
where $C^{\dagger}$ ($C$) is creation (annihilation) operator of
the cavity photon with frequency $\omega_{{\rm C}}$. Within the RWA and the dipole approximation, interaction of the dimer
with the electric field of the cavity mode is specified by the Hamiltonian
\begin{eqnarray} \label{CM}
H_{{\rm CM}}^{(j)} &  = & \frac{\Omega}{2}(CX^{\dagger}+C^{\dagger}X), \\
X^{\dagger} & = & |{\rm S}_{1}^{(j)}\rangle\langle {\rm g}^{(j)}|+\eta_{S}|{\rm S}_{n}^{(j)}\rangle\langle {\rm S}_{1}^{(j)}|+\eta_{T}|{\rm TT}_{n}^{(j)}\rangle\langle {\rm TT}^{(j)}|\nonumber
\end{eqnarray}
where $\Omega$ is the vacuum Rabi frequency for a single molecular
emitter. For $N$ identical molecules in the cavity, the Rabi splitting
in the Franck-Condon region can be approximated by the formula
\begin{equation}
\Delta_{R}=\sqrt{(\omega_{C}-\varepsilon_{S_{1}})^{2}+N\Omega^{2}}\label{Del}
\end{equation}
derived in the TC model (see, e.g., Ref. \cite{HTC}
and discussion therein). The state ${\rm S}_{1}^{(j)}$ is optically bright
from the ground state while the state ${\rm TT}^{(j)}$ is optically dark,
so only the ${\rm S}_{1}^{(j)}$ state is coupled to the cavity in Eq. (\ref{CM}).
Cavity-induced transitions between the lower-lying and upper-lying
states ${\rm S}_{1}^{(j)}$-${\rm S}_{n}^{(j)}$ and ${\rm TT}^{(j)}$-${\rm TT}_{n}^{(j)}$
are governed by the parameters $\eta_{S}$ and $\eta_{T}$. The total
Hamiltonian $H$ commutes with the number operator $N_{{\rm ex}}=C^{\dagger}C+\sum_{j,k}|k^{(j)}\rangle\langle k^{(j)}|$
and the total number of excitations is therefore conserved. The present SF model is constructed to investigate the nonlinear spectroscopic signatures in Section \ref{Spectroscopy}, in which the ground-state bleach (GSB), the stimulated emission (SE), and the excited-state absorption (ESA) are included.
It is worth mentioning that Hamiltonian (\ref{Htot}) is equivalent to the HTC model in the special case of $\lambda = 0$.

To explore the optimal parameters of a highly efficient cavity-controlled SF process, the higher-lying states excitation in Hamiltonian (\ref{HM}) can be neglected in the simulations for brevity. Thus, every rubrene molecule mediated by a CI could be modeled as a three-electronic-state (a ground state $\rm g$, an excited singlet $\rm S_1$ and a correlated triplet pair state $\rm ^1(TT)$)~\cite{sun1}.
Furthermore, in order to achieve a larger tunable model parameter space, for example, the higher average photon number state, the non-RWA terms should be taken into account in the cavity-molecule coupling Hamiltonian, which reads
\begin{eqnarray}\label{CM1}
H_{\rm CM}^{(j)}&=&\frac{\Omega^{(j)}}{2}(|{\rm g}^{(j)}\rangle\langle{\rm S_1}^{(j)}|+{\rm H.c.})(C^{\dagger}+C).
\end{eqnarray}
The non-RWA terms are particularly significant in physical experiments involving strong and ultrastrong coupling~\cite{Ann21}, short-time quantum dynamics~\cite{Zheng08}, high-precision spectral analysis~\cite{Li13}, as well as interactions between ultrashort laser pulses and matter~\cite{Lu16}. The non-RWA Hamiltonian (\ref{CM1}) is adopted to study the SF dynamics with different average photon numbers in Section \ref{dynamics}.

\begin{table}
\caption{Summary of bare SF material parameters and their optimal gain rates of TT yield $\tilde{\epsilon}_{\rm TT}/\epsilon_{\rm TT}$ in the cavity. where $\Delta\rm G$ is given by $\Delta\rm G=\omega_{\rm TT}-\omega_{\rm S_1}$, and $\epsilon_{\rm TT}$ ($\tilde{\epsilon}_{\rm TT}$) represents bared (polariton-assisted) TT yield, respectively.}
\begin{center}
\begin{tabular}{p{1.5cm}p{1.5cm}p{2.2cm}p{1.5cm}p{1.5cm}}
\hline
\hline
$\rm Molecule$ & $\Delta\rm G (meV)$ & $\epsilon_{\rm TT} (\%)$ & $\tilde{\epsilon}_{\rm TT}/\epsilon_{\rm TT}$ & Ref.  \\
\hline
$\rm Tetracene$ & $150$ & 0.01 & $\sim16$ & \cite{Mart}\\
$\rm Pentacene$ & $-110$ & 100  & $\sim1.4$ & \cite{Mart}\\
$\rm Hexacene$ & $-630$ & 48 & $\sim3.3$ & \cite{Mart}\\
$\rm Rubrene$ & $50$ & $17, 40, >99.9$ & $\sim 1.3$ & \cite{Takahashi,Maslennikov,Malin13}\\
\hline
\hline
\end{tabular}
\end{center}
\label{TT1}
\end{table}

The parameters characterizing several popular bare and cavity-manipulated   SF materials are listed in Table~\ref{TT1}.
As follows from this Table, incorporation of these materials in microcavities increases SF yield.
Hence integration of QED-cavities  into the design of photovoltaic devices offers a notable potential to enhance singlet-to-triplet conversion efficiency and  circumvent the Shockley-Queisser limit \cite{SQ,Markvart22}. However, the fulfilment of this program requires understanding of the microscopic mechanisms of the cavity-enhanced SF. In this context, it is essential  that the majority of the  SF  materials (for example, pentacene \cite{Musser,HGD20}, tetracene \cite{Herbert17,Xiao20}, and rubrene \cite{Miyata}) exhibit conical-intersection (CI) driven SF.

CIs are points of degeneracy in the potential energy surfaces (PESs) of isolated polyatomic molecules. Manifolds of these points of degeneracy, known as CI seams, undermine the Born-Oppenheimer approximation and open up effective and fast channels for the population and charge transfer~\cite{ConicalIntersections}.
In PPP  systems,  the situation is complicated by the fact that  CIs may be either intrinsic or cavity-induced \cite{Mukamel16,Kowalewski17}. In the strong coupling regime, an intrinsic molecular CI splits into a pair of polaritonic CIs~\cite{Gu2,sun1}. Furthermore, strong coupling to the cavity mode may produce so-called light-induced CIs \cite{Cederbaum12,Cederbaum13}, which occur even in  diatomic molecules possessing no intrinsic CIs~\cite{Cederbaum08,Csehi}. These light-induced CIs may be caused by the presence of additional (for example, rotational/orientational) DOFs.
Hence the coupling to the cavity creates multidimensional PES landscapes comprising cascades of intrinsic and cavity-induced/modified CIs~\cite{Sun6,Gu3,Gu2,sun1,Cho,Gu,Gu1,sun2}. On the one hand, this enormously complicates the situation and imposes high requirements on the numerical efficiency and accuracy of the theoretical methods used for the simulation of dynamic and spectroscopic responses of these systems. On the other hand, the cavity-induced/modified  CIs create additional pathways and shortcuts for the population  transfer, which can be used for engineering/optimizing the SF process~\cite{Gu3,sun1}.
 Yet, the polaritonic CIs in the strong-coupling regime may act as double-edged sword. For example, Gu {\it et al.}~uncovered that SF  can be significantly suppressed, since polaritonic CIs are shifted away from the Franck-Condon region in pentacene dimers~\cite{Gu2}. Hence enhancing the SF process in cavities remains a topic of contention.

\subsection{Polaritonic CI landscapes}
 In rubrene, the energy of the TT state, $\rm E_{TT}$, is slightly higher than the energy of the singlet state, $\rm E_{S_1}$.
 Hence SF -- without thermal activation \cite{Miyata} or excitation to higher-lying electronic states $\rm E_{S_n}$ \cite{Malin12} -- is rather inefficient. Furthermore, the usual mechanism of polaritonic enhancement of SF through the cavity-finetuned resonance of the LP and TT states, is ineffective in rubrene, too~\cite{Ribeiro2018,Takahashi,Mart}. Hence rubrene is  a suitable system for exploring  novel cavity-induced  mechanisms of the SF enhancement.

\begin{figure*}[tbp]
\begin{minipage}[t]{0.4\linewidth}
\centering
\includegraphics[scale=0.45,trim=0 0 0 0]{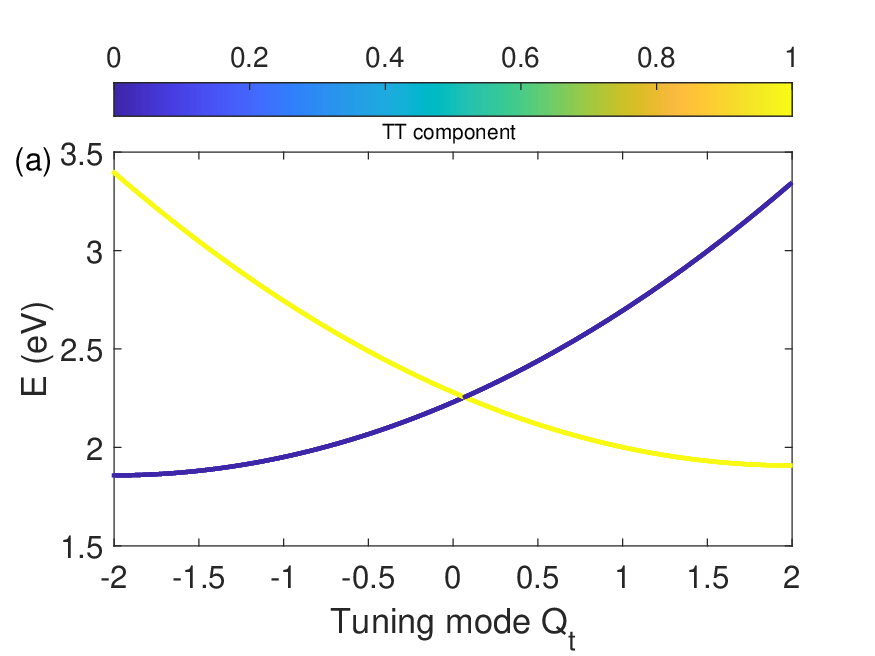}
\end{minipage}
\begin{minipage}[t]{0.4\linewidth}
\centering
\includegraphics[scale=0.45,trim=0 0 0 0]{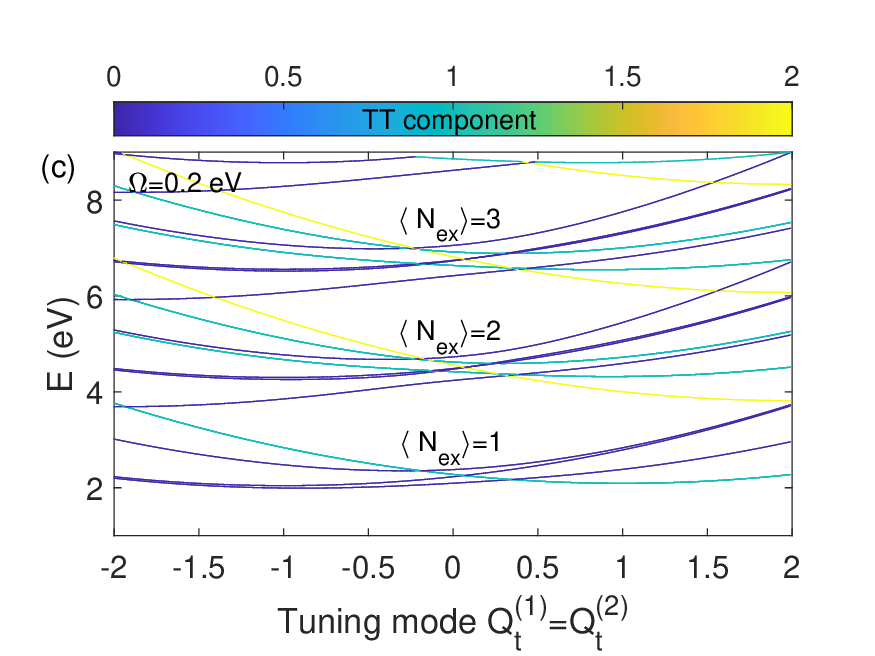}\\
\end{minipage}
\begin{minipage}[t]{0.4\linewidth}
\centering
\includegraphics[scale=0.45,trim=0 0 0 0]{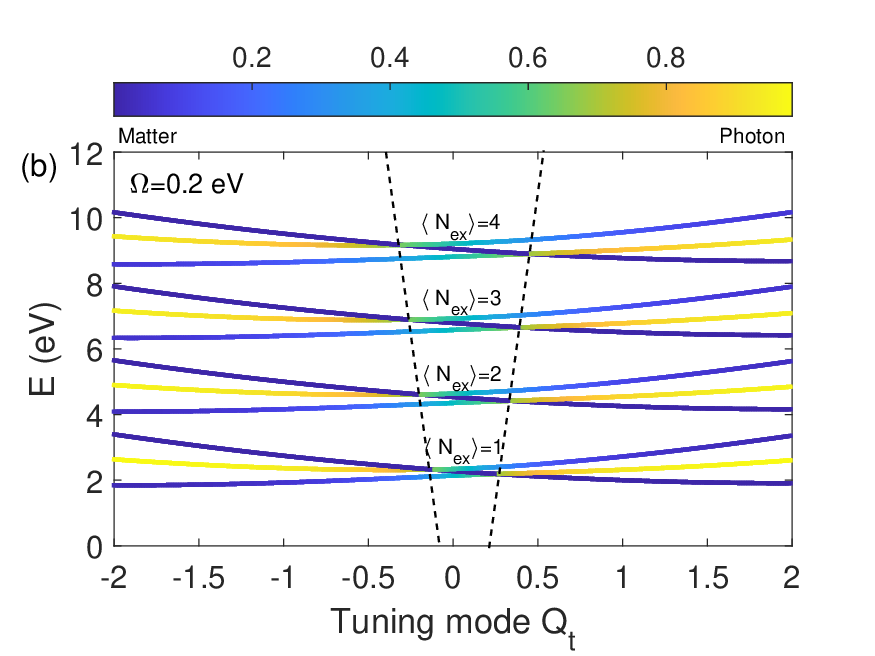}
\end{minipage}
\begin{minipage}[t]{0.4\linewidth}
\centering
\includegraphics[scale=0.45,trim=0 0 0 0]{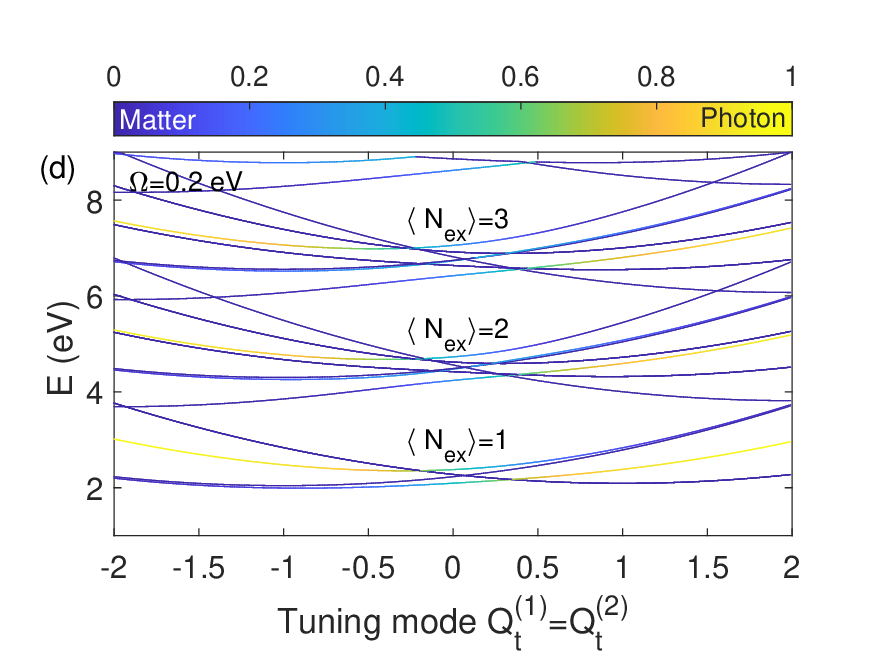}
\end{minipage}
\caption{Cuts of the adiabatic PESs along the tuning mode ($Q^{(j)}_c=0$) for one (left column; panels a, b) and two (right column; panels c, d) RDs in the cavity for the first few polaritonic manifolds $N$. (a) No coupling to the cavity mode ($\Omega=0$), the bare CI is located in the Franck-Condon region at  $Q^{(1)}_t=0.07,E=2.256~\rm eV$.
(b, c, d) Strong coupling to the cavity mode ($\Omega=0.2~\rm eV$).
The fraction of the TT states ($W_{\rm TT}$ of Eq. (\ref{TT})) is indicated through the line color coding of PESs (a, c), while fraction of the photonic states ($W_{\rm ph}$ of Eq. (\ref{W})), is indicated through  the line color coding of the PESs (b, d).
The dash lines in panel (b) show positions of the polaritonic CIs.  Adapted from \cite{sun1}. Copyright American Chemical Society.}
\label{Fig1}
\end{figure*}

Following Ref.~\cite{sun1}, we consider a microcavity with one or two RDs placed into it. {To focus on the net effect of the cavity-induced control mechanisms, the loss of photons in the cavity is neglected. This approximation is reasonable. For example, ultrahigh-Q PhCnB cavities with a quality factor of approximately $\rm Q\sim750,000$ have been fabricated through the utilization of a five-hole taper design \cite{Deotare}. Moreover, the nanobeam cavity has been optimized to ensure a quality factor of 1700 and a transmission efficiency of $90\%$. When integrated with the bowtie, the hybrid photonic-plasmonic cavity attains a quality factor of 800 and a transmission rate of $20\%$ \cite{Conteduca}. Thus, the timescale of the studied SF processes (hundreds of femtoseconds) is short in comparison with the lifetimes of such cavities.}
In this setup, the $j$th  RD  participating in the CI-mediated SF is  modeled as a system with three electronic states (the ground state ${\rm S}_0^{(j)}$, the singlet excited state ${\rm S}_1^{(j)}$, and the correlated triplet pair state ${\rm TT}^{(j)}$) and two vibrational modes (the tuning mode with a frequency $\Omega_t = 0.186$ eV and coordinate $Q^{(j)}_t$, and the coupling mode with a frequency $\Omega_c  = 0.0154$ eV and coordinate $Q^{(j)}_c$)~\cite{Miyata,sun1}. The electronic energies of the singlet and triplet  states are fixed at  $E_{{\rm S}_1}^{(j)}=2.23$ eV and $E_{{\rm TT}_1}^{(j)}=2.28$ eV.
The cavity is modelled via a single photonic mode of a frequency $\omega_c=2.256$ eV, which is nearly in the electronic resonance with the  ${\rm S}_1^{(j)}$ and ${\rm TT}_1^{(j)}$ states. $\Omega$ determines  the coupling of the RDs to the cavity mode. If not explicitly stated otherwise, we set  $\Omega=0.2$ eV, which corresponds to the strong cavity-RD coupling regime. To increase space of the adjustable  parameters, the RWA in the description of the cavity-matter interaction was not applied.

 All levels of the combined RD(s) + cavity system can be classified according to the expectation value of the number operator
 \begin{equation}\label{Nex}
  N=\langle \Psi_n({\bf Q})| \hat{N}_{\rm ex}|\Psi_n({\bf Q})\rangle
 \end{equation}
 where $|\Psi_n({\bf Q})\rangle$ is the adiabatic eigenfunction of the  polaritonic Hamiltonian
(since the cavity mode is in electronic resonance with the bright RD state, the RWA is almost exact and the number of excitation is nearly conserved).
$N$ specifies the excitation manifold of the polaritonic PESs. For example, $N=0$ corresponds to  the collective ground state, $N=1$ yields the singly-excited manifold, etc. To characterize and visualize various states, it is  elucidating to introduce several quantifiers. For instance,
\begin{eqnarray}\label{W}
W_{\rm ph}=1-\sum_{\boldsymbol{k}}\sum_{n_c \ne 0}|\langle \boldsymbol{k}, n_c|\Psi_n({\bf Q})\rangle|^2
\end{eqnarray}
gives a fraction of the pure photonic excitation in a certain manifold $N$. The sum in Eq. (\ref{W}) runs over the  components $k_j= {\rm S}^{(j)}_1,{\rm TT}^{(j)}$ of vector $\boldsymbol{k}$. The number of these  components  matches  the number of RDs in the cavity, and $n_c$ is the quantum number specifying the photonic-mode excitation.
Similarly, the TT component of the polaritonic state can be defined as
\begin{eqnarray}\label{TT}
W_{\rm TT}=\sum_{\boldsymbol{k} \in {\rm TT}}\sum_{n_c}|\langle \boldsymbol{k}, n_c|\Psi_n({\bf Q})\rangle|^2,
\end{eqnarray}
where the notation $\boldsymbol{k} \in {\rm TT}$ means that at least one of the components $k_j$   must contain ${\rm TT}^{(j)}$.

The fractions of the TT states ($W_{\rm TT}$ of Eq. (\ref{TT})) and  photonic states ($W_{\rm ph}$ of Eq. (\ref{W})) give a convenient visualization of the polaritonic PESs, which is implemented through the line color coding in Figure~\ref{Fig1}.
Panels (a, b)  represent a single RD in the cavity, and show the cut of the polaritonic PESs along the symmetric tuning mode $Q_t$  at $Q_c=0$. Panel (a) corresponds to a cavity-free RD. Hence, ${\rm S}_1^{(1)}$ and ${\rm TT}^{(j)}$
are not dressed with the cavity mode, and only the intrinsic RD CI at $Q_t=0.07,E=2.256~\rm eV$ exists in this case. It follows that this CI located in the Franck-Condon region provides the only SF pathway. In the presence of cavity, the PESs change substantially, and cascades of polaritonic CIs appear in Figure~\ref{Fig1}(b). As the coupling to the cavity increases, the two polaritonic CIs move away from the Franck-Condon point, the distance between the pair of polaritonic CIs increases with $N$, and positions of these CIs lie along the V-shaped dash lines in Figure~\ref{Fig1}(b). {A very similar situation takes place  in the  pentacene dimer, in which the polaritonic CI is pushed away from the Franck-Condon region,  suppressing the SF process~\cite{Gu2}.} However, it is essential that the fraction of the hybrid matter-photon PESs increases with $N$ near the Franck-Condon point. This reversely facilitates the cavity-assisted SF.

The cuts of PESs for two RDs strongly coupled to the cavity mode are shown in Figures~\ref{Fig1}(c) (highlighting the TT character of the PESs) and (d) (highlighting the photonic character of the PESs). First of all the number of the coupled PESs apparently increases with the number of RDs in the cavity. If for a single RD  we have 3 PESs for $N\ge 1$, for two RDs there are 5 PESs for $N = 1$, and 9 PESs for $N\ge 2$. The PES with the CIs at $Q^{(1)}_t = Q^{(2)}_t= 0.07$ eV, which corresponds to both RDs in the triplet state, exists for all $N$. With the exception of this PES, the cavity-RD coupling repels polaritonic CIs from the central region, but the effect is not so pronounced as in the case of a single RD.

\subsection{SF dynamics Mediated by Polaritonic CIs}\label{dynamics}

Let us explore how the polaritonic PESs coupled through cascades of the cavity-induced CIs affect the SF dynamics.
{Vendrell suggested the number of photons in the cavity, $N$, can be used as a ``control knob", since
stimulated emission induced by a cavity resonantly coupled to a specific molecular reaction coordinate strongly depends on $N$~\cite{Vendrell1}.} To achieve this, it is
worthwhile to study the impact of the initial excitation of the photonic mode on the ensuing singlet-to-triplet population transfer. In our simulations, the cavity mode is assumed to be initially, at $t=0$, in the coherent state
\begin{eqnarray}
|\mu_{1}(t)\rangle = \exp(-|\mu_1|^2/2)\sum_{n_c=0}^{\infty}\frac{\mu_1^{n_c}}{\sqrt{n_c!}}|n_c\rangle
\end{eqnarray}
where $\mu_1$ is the displacement parameter. For a specific $n_c$, the main contribution to the coherent state gives the term with $\mu_{1}(0)=\sqrt{n_c}$. Hence we will systematically study how the initial preparation of the RD system in the manifold $N$, where the population is initially placed in the singlet state and the cavity mode is in the coherent state with $\mu_{1}(0)=\sqrt{N}$ affects SF in the regime of strong RD-cavity coupling.
This is illustrated by Figure~\ref{Fig2} for one (a, c) and two (b, d) RDs in the cavity.

\begin{figure*}[tbp]
\centering
\includegraphics[scale=0.4,trim=50 0 40 0]{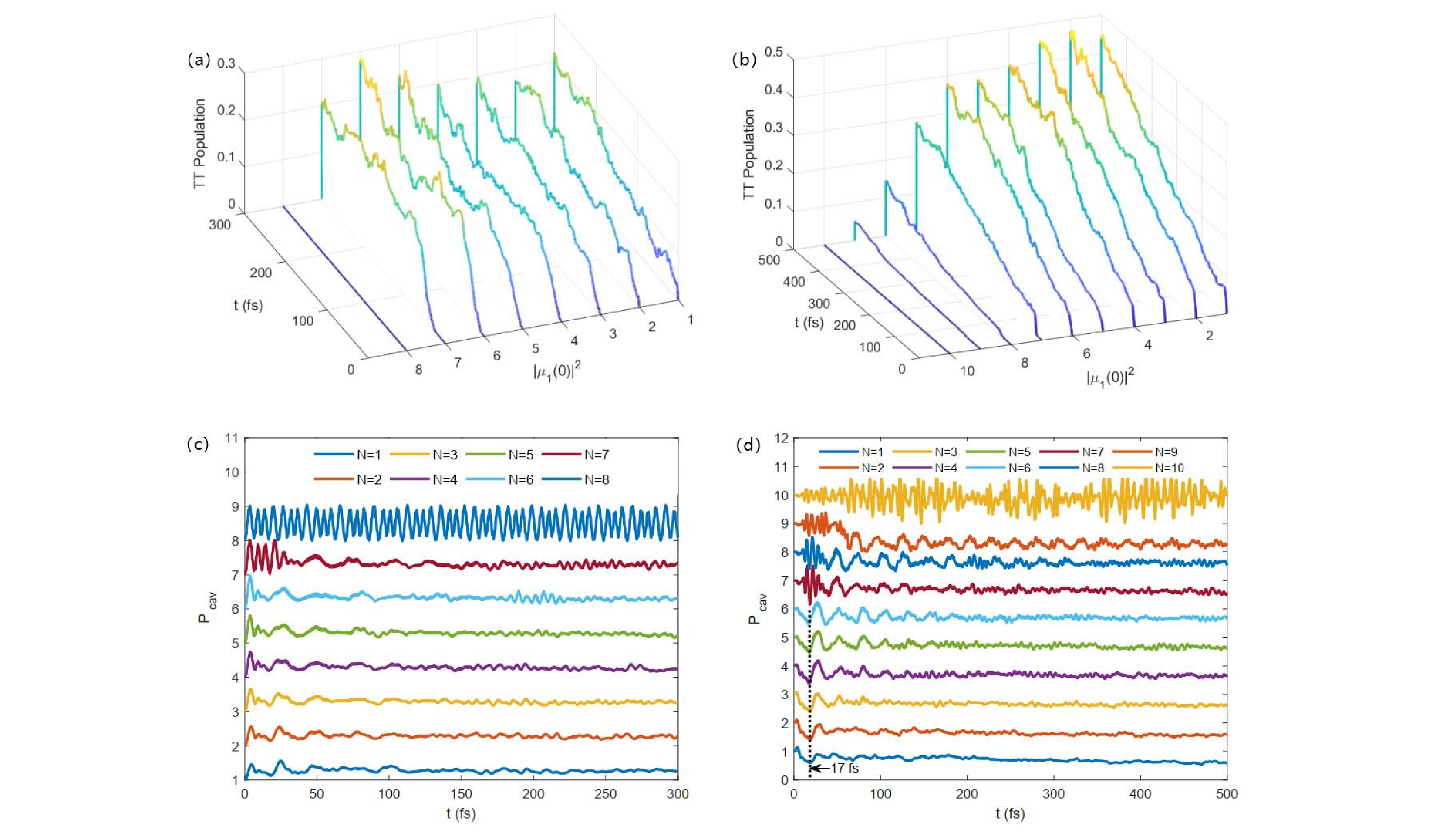}
\caption{Population dynamics of the triplet state (upper panels) and average photon mode populations $P_{\rm cav}(t)$ (lower panels)  for several initial photon-mode  pumping numbers $|\mu_1(0)|^2=N$.
Left column (panels a, c): one  RD. Right column (panels b, d): two RDs. Adapted from \cite{sun1}. Copyright American Chemical Society.}
\label{Fig2}
\end{figure*}

Let us begin with the considerations of the upper panels, which
show the triplet populations for different $N$.
For a single RD, the triplet population  -- on the average -- increases  with $N$ up to $N=6$. For example, the population of   $\approx 0.28$ is observed at 300 fs for $N=6$, which is approximately twice larger than the population obtained in the same model without a cavity.
As long as these photons remain in the cavity and are available for the next ``$\rm S_0\to S_1\to TT \to charge~transfer$'' cycle, the SF efficiency will increase.
However, the SF efficiency starts to decrease at $N=7$ and is fully suppressed for $N\ge8$.
This effect can rationalized from the different points of view.
On the one hand, it can be viewed as excitation
localization on the singlet state ${\rm S}_1^{(1)}$ and on the photonic mode, which renders the ${\rm TT}^{(1)}$ effectively decoupled and the SF  switched off (see Refs. \cite{Localization19,Localization20} for engineering of long-lived localized states in the cavity-QED systems).  A qualitatively similar localization is realized, for example, in the spin-boson model in the regime of the strong coupling of the spin system to the bosonic bath \cite{Wu}, even at different system preparations \cite{LP23}.
Alternatively, the effect can be envisaged as an effective annihilation of the nonadiabatic coupling between the singlet and triplet states  \cite{Gelin09}. It operates through the combined effect of the cavity field (through Rabi cycling) the system itself (through the intrinsic ${\rm S}_1^{(1)}-{\rm TT}^{(1)}$ CI), since the photonic mode does not directly couple the singlet and triplet states.
Note also that the population evolution of the triplet states starts not from $t=0$, but from a characteristic time $\tau_N$ which increases with $N$. This indicates that the wavepacket needs $\approx \tau_N$ to reach the polaritonic CIs.

A qualitatively similar situation takes place for 2 RDs in the cavity (Fig.~\ref{Fig2}(b)),  with a notable modifications though.  For $N=1$, the total triplet population reaches  $\approx 0.45$ at $t=500$ fs, which is $\approx 1.5$ higher than for  $N=0$. As demonstrated in the previous section (Figs.~\ref{Fig1}(c, d)) photonic mode helps to generate multiple polaritonic CIs in the Franck-Condon, owing  to the elevated number of the coupled cavity-induced CIs and existence of the doubly-excited states in the manifolds $N$. This cooperativity facilitates the population transfer, inaugurating the onset of the cavity-assisted SF.
A comparable relatively high total triplet population is obtained with $N=2$ (the largest) and with $N=3$, which is attributed to similar polaritonic PESs for $N=1,\,2, \, 3$.
As shown in Fig.~\ref{Fig1}(c, d), elevated $N$   do not substantially shift positions of the coupled CIs in the Franck-Condon region.  Therefore, at variance with the case of a single RD, no SF enhancement with $N$ is essentially observed. On the contrary, as the photon displacement increases, the ${\rm TT}^{(1)}$ decoupling  becomes more pronounced, which results in decreasing triplet populations and, eventually,  a full SF switching-off at $N=10$. This triplet population decrease can be attributed to the influence of the central  polaritonic CI (Fig.~\ref{Fig1}(c, d)) which does not exist in the single-RD system (Fig.~\ref{Fig1}(a, b)).

\begin{figure*}[tbp]
\centering
\includegraphics[scale=0.4]{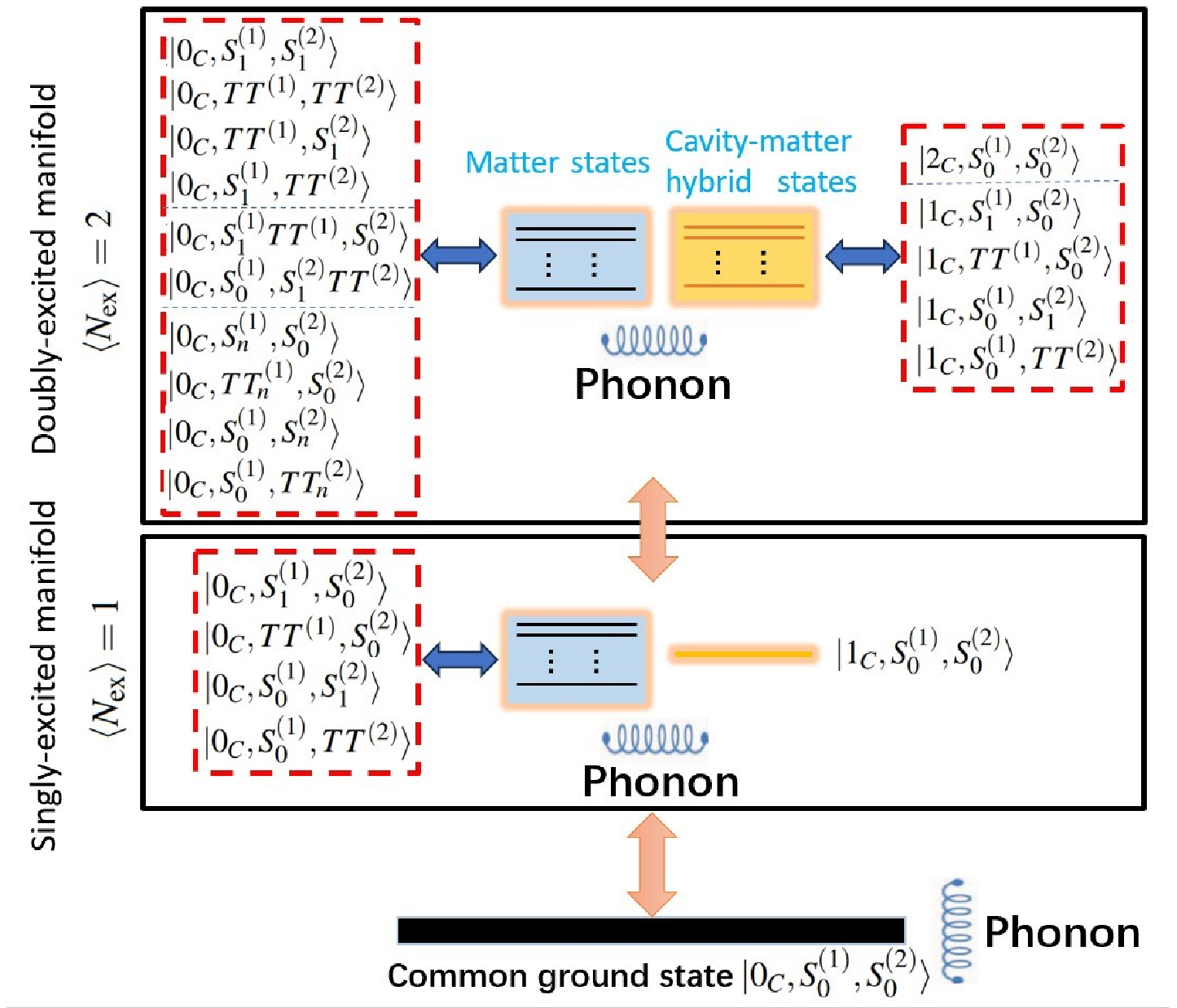}
\caption{Energy levels of two rubrene dimers in the cavity. The superscripts $(1)$ and $(2)$ number the dimers. }
\label{Levlels}
\end{figure*}

Let us now inspect evolutions $P_{\rm cav}(t)$ of the cavity-mode populations vs. the initial photon pumping numbers $N$ ($P_{\rm cav}(0)=N$) which deliver a complementary information.
$P_{\rm cav}(t)$ for a single RD is depicted in Fig.~\ref{Fig2}(c). For $N$ from 1 to 6, the populations exhibit similar  behavior: During the first 50 fs, they show several oscillations with a rough period of 22 fs which is related to the tuning mode. Then they reach a quasi-steady-state $\langle P_{\rm cav} \rangle$ value at longer times. It is  worth mentioning that $\langle P_{\rm cav} \rangle$ are  slightly larger than  the initial populations $P_{\rm cav}(0)=N$. It reveals the population redistribution due to the coupling of the cavity  mode and the ${\rm S}_1^{(1)}$ state. $P_{\rm cav}(t)$ for $N = 7$ shows high-amplitude oscillations during the first $\sim 40$ fs, which matches the time $\tau_7$ during which the corresponding triplet population  in Fig.~\ref{Fig2}(a) is almost zero.   For $N = 8$, the high-amplitude oscillations persist over the entire evolution of $P_{\rm cav}(t)$, which is a signature of the localization of the population on the photonic mode and the singlet state, which results in the SF stopping.
Interestingly, the period of these fast oscillations reveals  the effective Rabi frequency $\Omega_N\approx\Omega \sqrt{N}$ ($2\pi/\Omega_8\approx 7.3$ fs), which is defined under the assumption that the effective coupling to the cavity  scales  $\sim \sqrt{N}$.

Fig.~\ref{Fig2}(d) is a counterpart of Fig.~\ref{Fig2}(c), but for two RDs coupled to the cavity.  For $N$ from 1 to 6, $P_{\rm cav}(t)$ reach their first local minima at $t \approx  17$  fs (dashed line in panel d), which indicates that a fraction of the photonic-mode population is transferred to the doubly excited electronic states of the RDs. Then, $P_{\rm cav}(t)$  exhibit oscillations with a period of $\sim 26$ fs, which cannot be directly related to the model parameters. In general, excitation to  $1\le N \le 6$ reflects population transfer between photonic mode  and excited states of the RDs.  For $7 \le N \le 9$, $P_{\rm cav}(t)$ shows fast Rabi oscillations at the initial stage, which correlate with the absence of the triplet excitation at the same time interval (cf.~Fig.~\ref{Fig2}(b)). For $N = 10$, fast Rabi oscillations spread over the entire $P_{\rm cav}(t)$ evolution, which is the signature of the SF switching-off in Fig.~\ref{Fig2}(b).
In the long-time limit, the $P_{\rm cav}(t)$  reach quasi-steady states $\langle P_{\rm cav} \rangle$ the values of which  are lower than the initial values $P_{\rm cav}(0)$. This indicates a net population transfer from the photonic mode to the doubly-excited electronic states of RDs.

The above analyses demonstrate that the SF efficiency is primarily determined by two factors. The first factor is the position of the polaritonic CIs (see Figure~\ref{Fig1}). If the CIs are located in the Franck-Condon region, the subsequent population transfer is facilitated and the higher SF efficiency is achieved. The second factor is the initial wavepacket preparation and the state of the cavity which is controlled by the photonic-mode pumping number $|\mu_1(0)|^2 = N$.
This later parameter plays the crucial role, increasing the SF efficiency for small $N$ and fully suppressing SF once $N$ reaches a certain critical value. {Similar results are obtained in Ref.~\cite{Vendrell1}. If the number of photons is higher than the number of the  available excited states, the effective cavity-matter coupling may increase as well, leading to slower motion of the nuclear wave packet, which gets partially trapped in the region where the molecular electronic energy gap becomes resonant with the cavity.}

{Furthermore, collective effects, such as polaron decoupling and collective dark states, have significant impacts on the cooperative polariton dynamics~\cite{Gu2}. SF dynamical pathways can also be modified by strong photon-matter coupling. For instance, the SF channel via the UP state impedes vibronic losses to the dark-states manifold, which evidently enhances the SF efficiency~\cite{Vendrell3}.}

\subsection{Spectroscopic signatures of Polaritonic CIs}\label{Spectroscopy}

As aforementioned, strong coupling of RDs to the cavity mode drastically alters the SF efficiency owing to the cohesive action of the cavity-induced and intrinsic CIs. This processes are imprinted into nonlinear femtosecond spectroscopic responses of these systems.
To find spectroscopic signatures of the polaritonic SF  occurring in multidimensional  PES landscapes, the signals must be accurately simulated and  adequately interpreted.
For this to achieve, it is essential to realistically model RDs. Indeed, all third-order spectroscopic responses consist of three contributions, the GSB, the SE, and the ESA \cite{Darius04,Leonas}. The GSB and SE reflect the photoinduced processes in the manifolds with $N=0$ (the ground state) and $N=1$ (singly-excited states), while ESA involves also the doubly-excited  states, $N=2$. Hence modeling individual RDs  as systems containing only singly-excited states ${\rm S}_1^{(j)}$ and ${\rm TT}^{(j)}$ is insufficient: we have to include higher-excited states ${\rm S}_n^{(j)}$ and ${\rm TT}_n^{(j)}$ \cite{sun2}.
 Following \cite{Miyata,sun2}, we set $E_{{\rm S}_n}^{(j)}=4.33$ eV and $E_{{\rm TT}_n}^{(j)}=4.68$ eV in our simulations.
For two RDs in the cavity, we thus obtain a single ground state,  5 singly-excited states and 15  doubly-excited states which are
specified in Fig. \ref{Levlels}. The states, in turn, can be subdivided into the matter states ($n_c=0$, left part of  Fig. \ref{Levlels}) and hybrid states ($n_c>0$, right part of  Fig. \ref{Levlels}).

\begin{figure*}[tbp]
	\begin{minipage}[t]{0.3\linewidth}
		\centering
		\includegraphics[scale=0.39,trim=0 0 0 0]{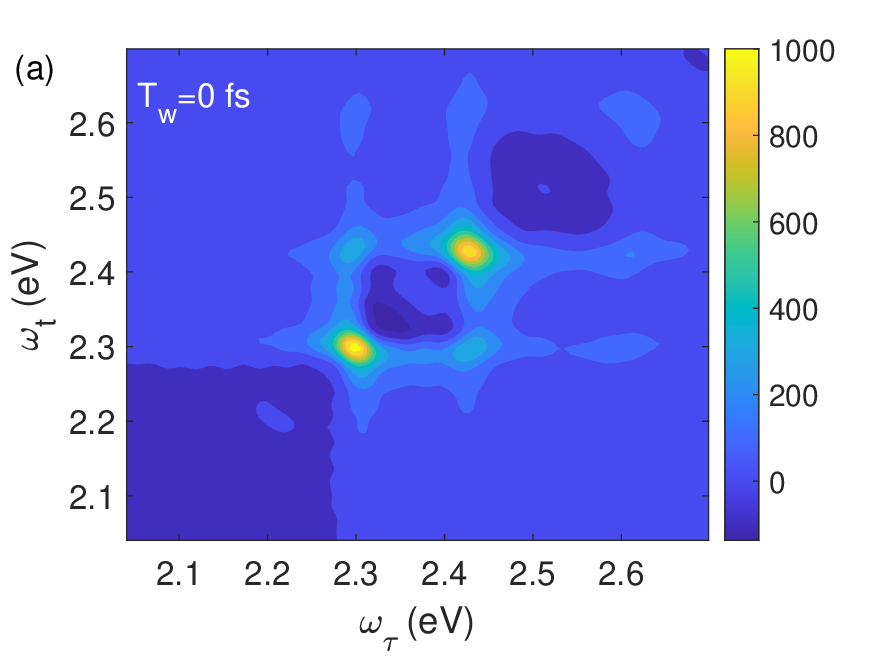}
	\end{minipage}
	\begin{minipage}[t]{0.3\linewidth}
		\centering
		\includegraphics[scale=0.39,trim=0 0 0 0]{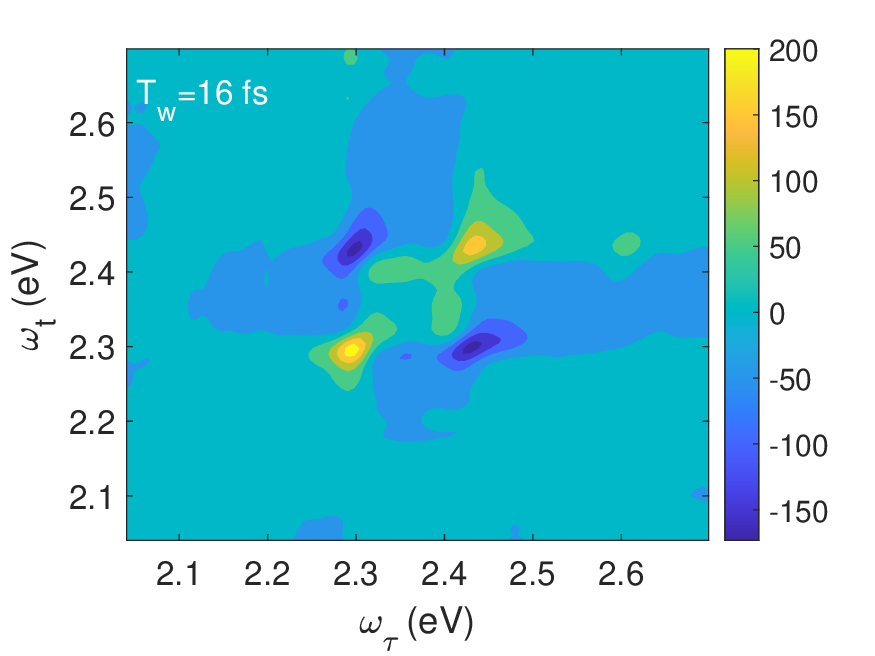}
	\end{minipage}
	\begin{minipage}[t]{0.3\linewidth}
		\centering
		\includegraphics[scale=0.39,trim=0 0 0 0]{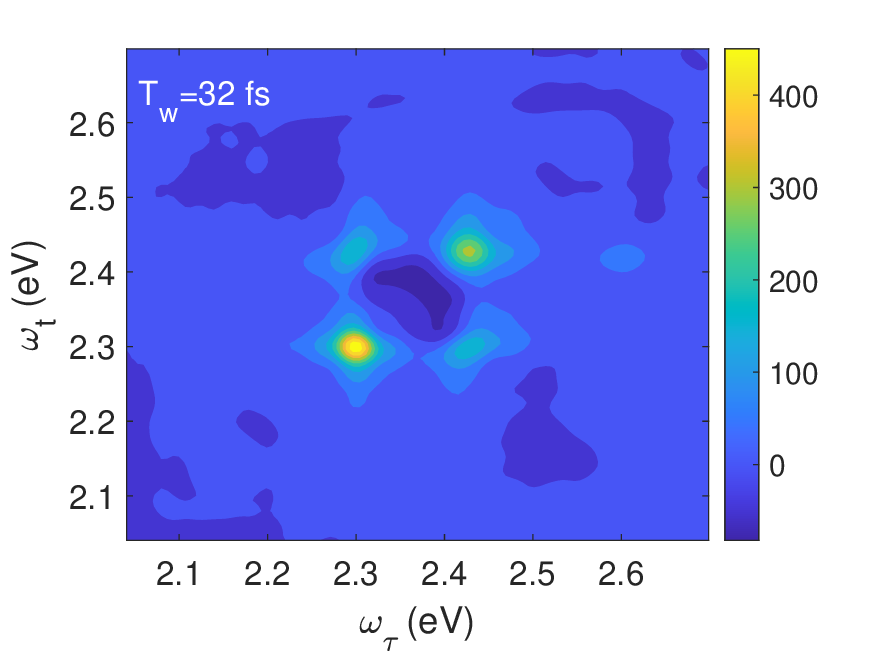}
	\end{minipage}
	\begin{minipage}[t]{0.3\linewidth}
		\centering
		\includegraphics[scale=0.39,trim=0 0 0 5]{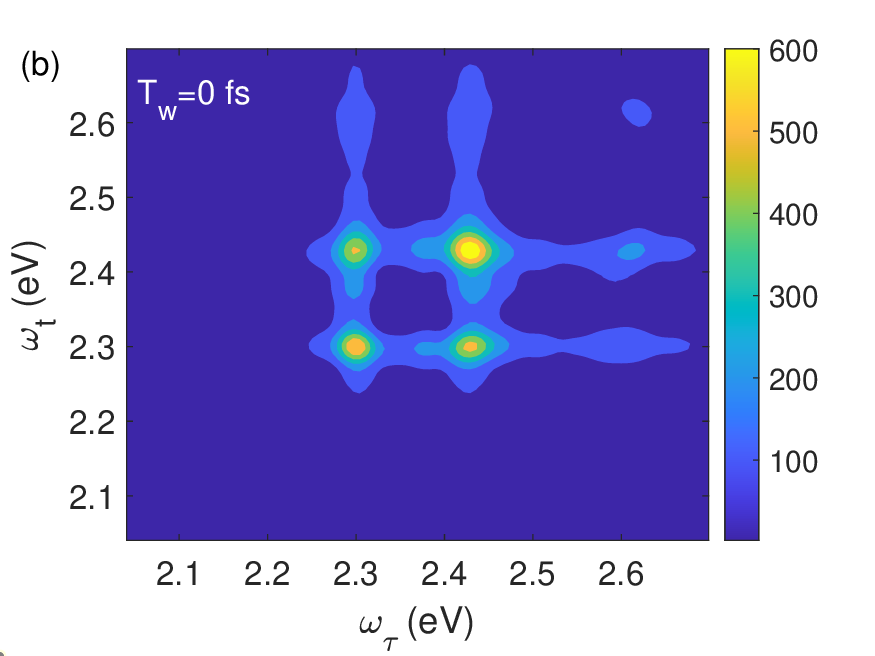}
	\end{minipage}
	\begin{minipage}[t]{0.3\linewidth}
		\centering
		\includegraphics[scale=0.39,trim=0 0 0 5]{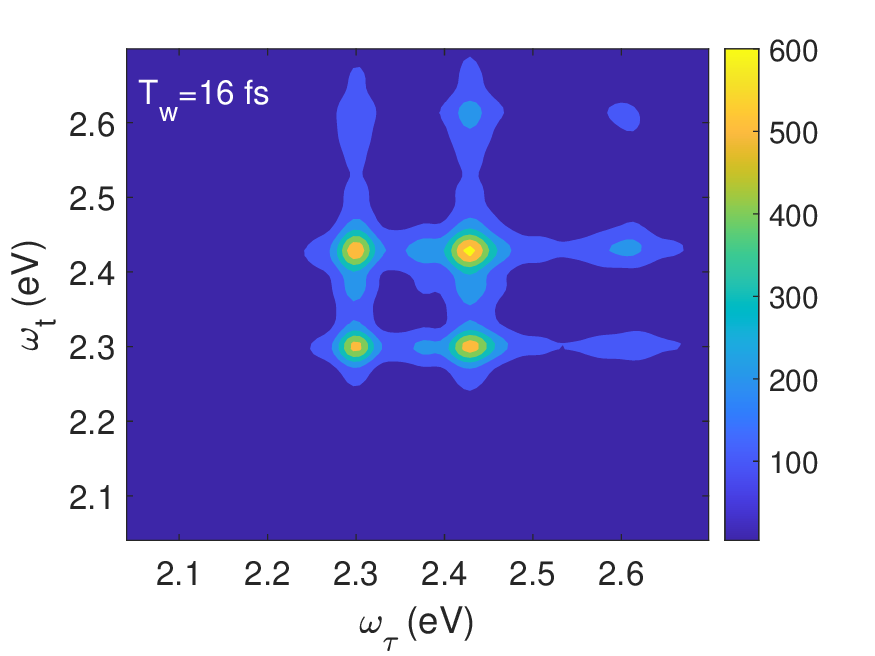}
	\end{minipage}
	\begin{minipage}[t]{0.3\linewidth}
		\centering
		\includegraphics[scale=0.39,trim=0 0 0 5]{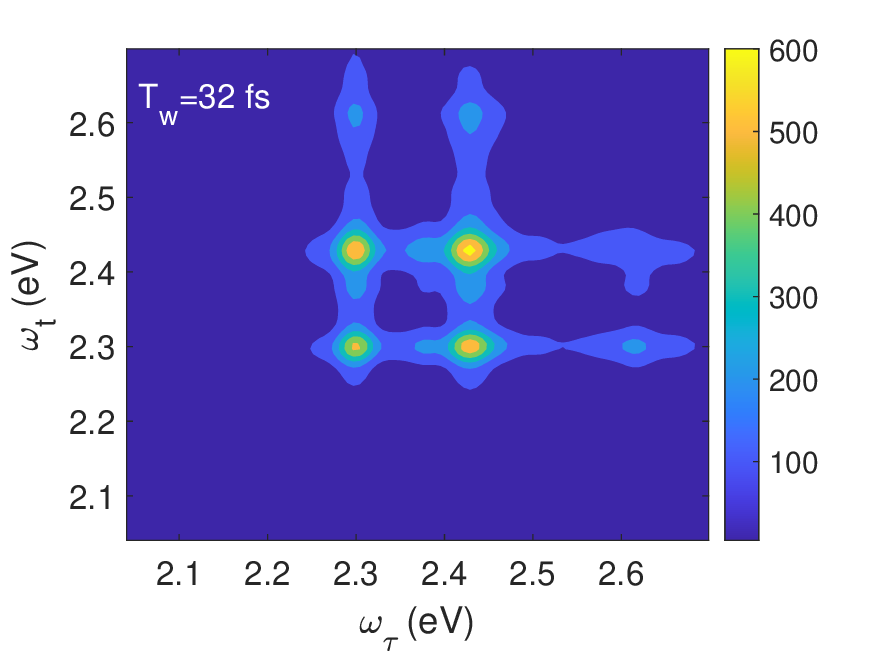}
	\end{minipage}
	\begin{minipage}[t]{0.3\linewidth}
		\centering
		\includegraphics[scale=0.39,trim=0 0 0 5]{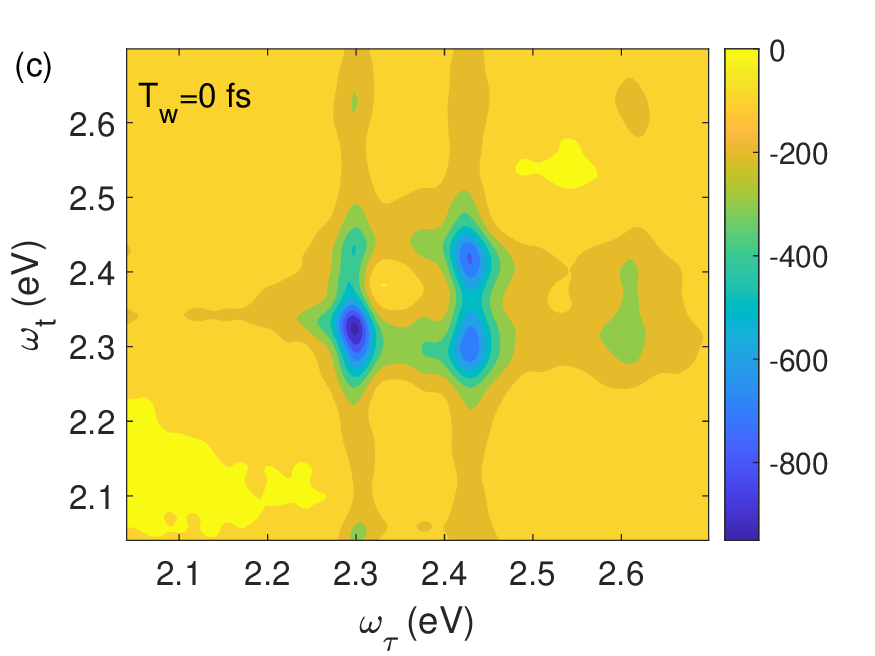}
	\end{minipage}
	\begin{minipage}[t]{0.3\linewidth}
		\centering
		\includegraphics[scale=0.39,trim=0 0 0 5]{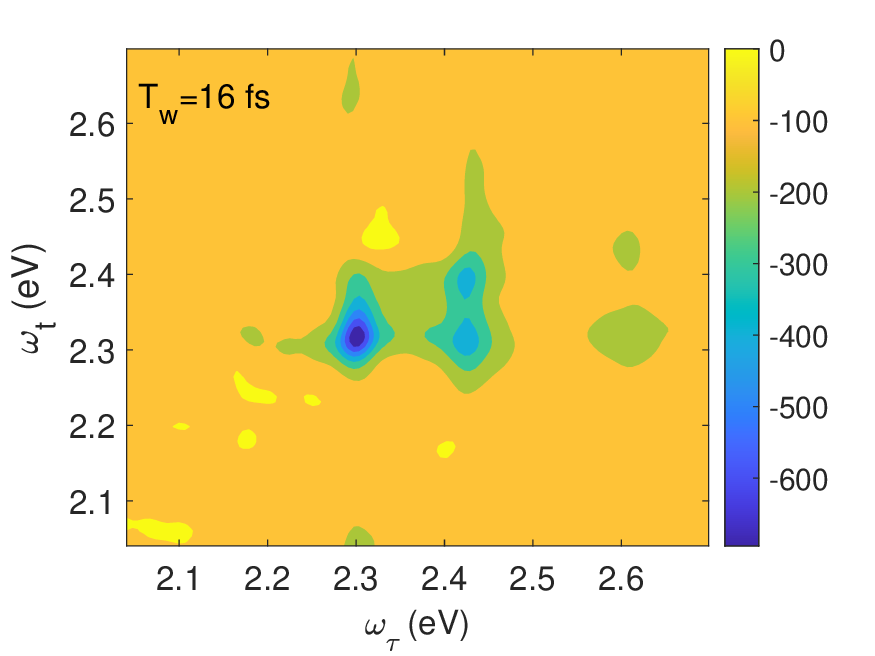}
	\end{minipage}
	\begin{minipage}[t]{0.3\linewidth}
		\centering
		\includegraphics[scale=0.39,trim=0 0 0 5]{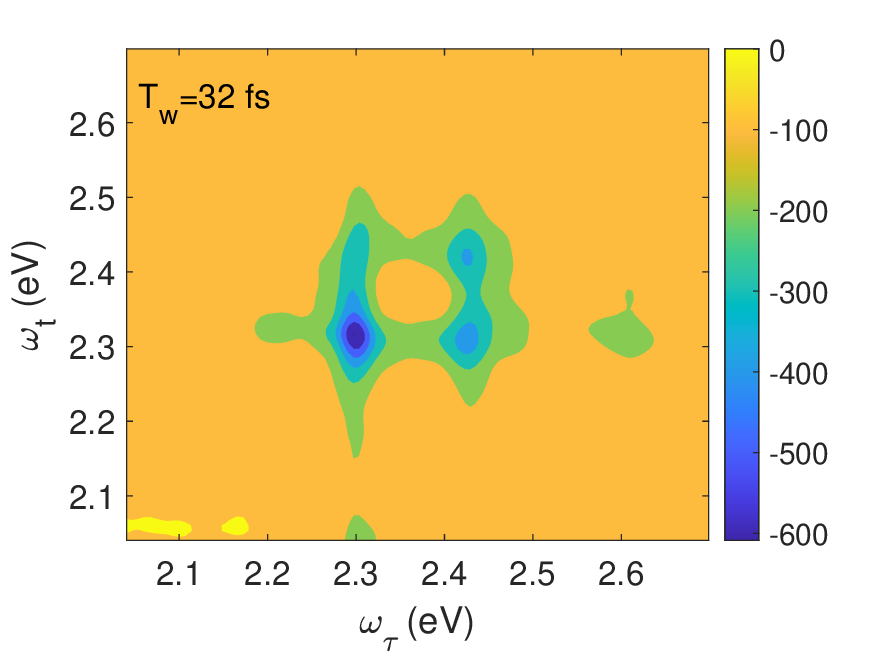}
	\end{minipage}
	\begin{minipage}[t]{0.3\linewidth}
		\centering
		\includegraphics[scale=0.39,trim=0 0 0 5]{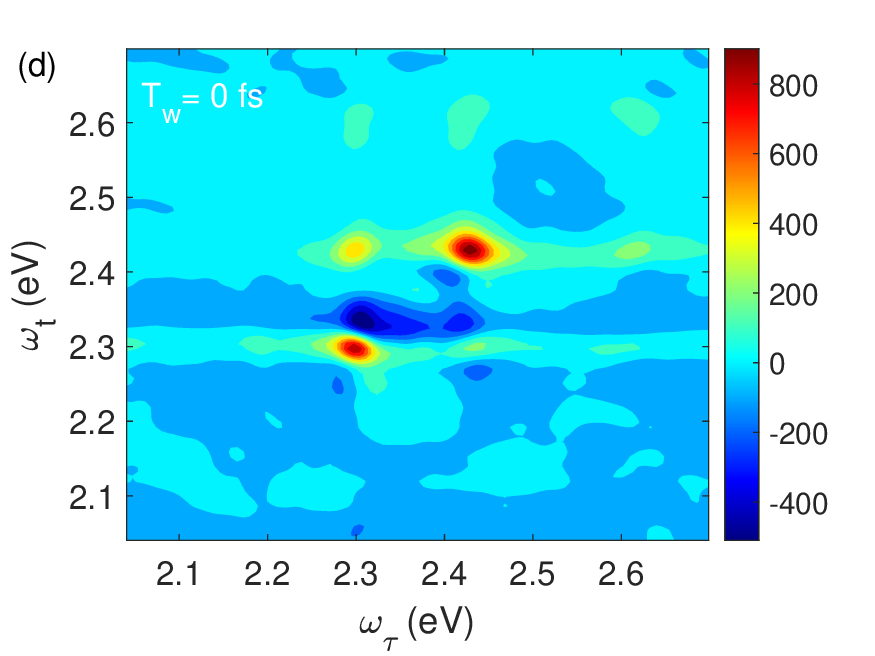}
	\end{minipage}
	\begin{minipage}[t]{0.3\linewidth}
		\centering
		\includegraphics[scale=0.39,trim=0 0 0 5]{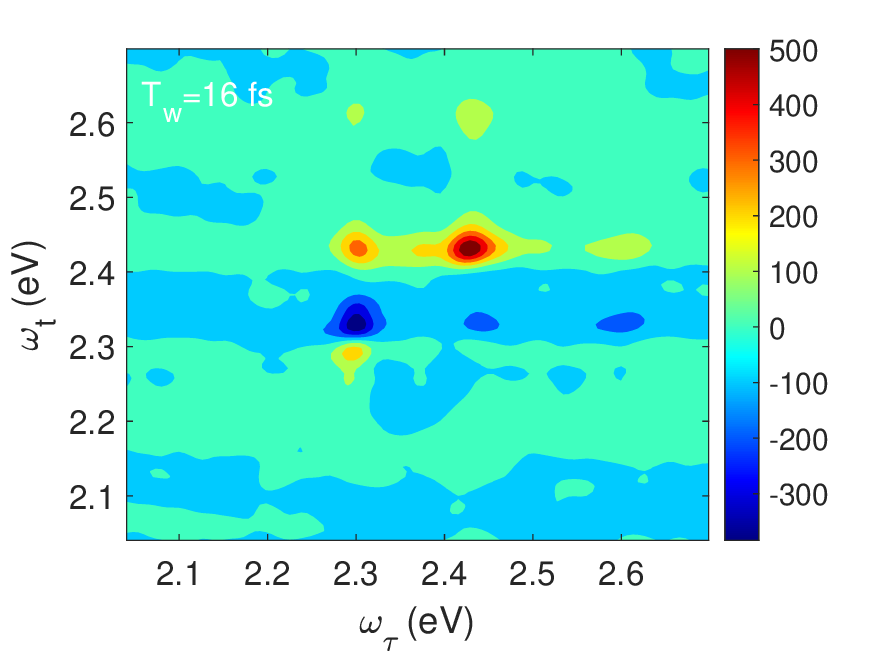}
	\end{minipage}
	\begin{minipage}[t]{0.3\linewidth}
		\centering
		\includegraphics[scale=0.39,trim=0 0 0 5]{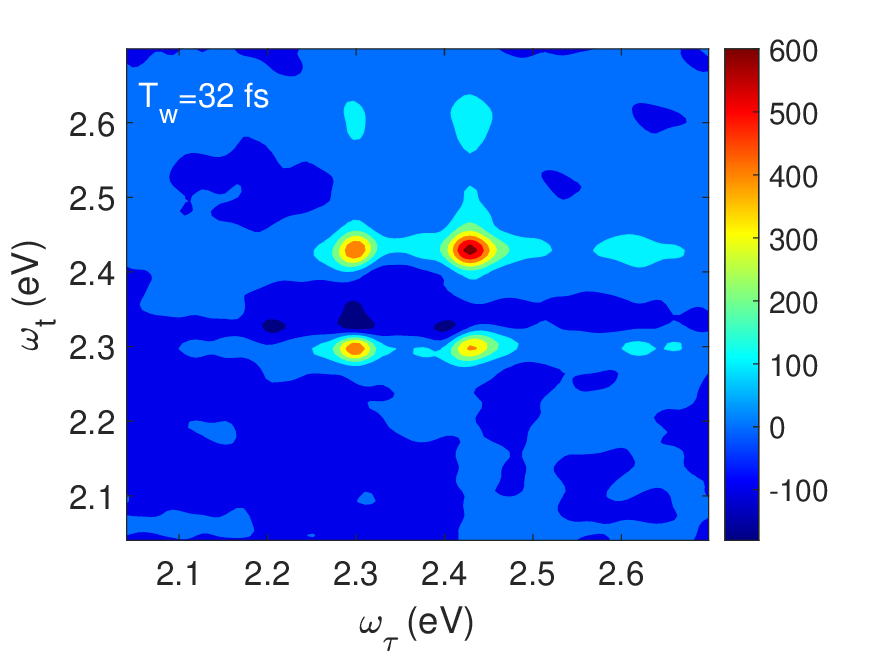}
	\end{minipage}
	\caption{Real part of the SE (row (a)), GSB (row (b)), ESA (row (c)) contributions
		and the total 2DES spectrum (row (d)) of the two RD system  for several
		population times $T_{w}$ indicated in the panels. The spectra are
		computed in the impulsive limit. Adapted from \cite{sun2}. Copyright American Chemical Society.}
	\label{Fig4}
\end{figure*}

Here we consider electronic two-dimensional (2D) spectra,  because  electronic 2D spectroscopy (2DES) is one of the  most powerful and information-rich third-order   spectroscopic techniques (see \cite{Scholes22} and references therein). Qualitatively, 2DES spectra
 $S(\omega_{\tau},T_{w},\omega_{t})$ can be envisaged as a $T_w$-dependent 2D $(\omega_{\tau},\omega_{t})$ maps, which reveal correlations between the states of the system excited at a frequency $\omega_{\tau}$ at time zero and probed at a frequency $\omega_t$ at time $T_w$.
 2DES spectra $S(\omega_{\tau},T_{w},\omega_{t})$ simulated for the two RD system strongly coupled to the cavity mode  are presented in Fig.~\ref{Fig4} for different population times $T_{w}$. From top to bottom, shown are the SE (row (a)),  GSB (row (b)), and ESA (row (c)) contributions, as well as the total signal (row (d)). The reader is referred to Appendix \ref{md2} for more derivation details on the various nonlinear response functions within the framework of the multi-D$_2$ Ansatz.

The SE contribution (row (a)) possesses two pronounced diagonal peaks
(DPs) and two cross-peaks (CPs) (similar situation takes place in the purely polaritonic dimer system \cite{Bondarev13}). The DP1 ($\omega_{\tau}=2.297~{\rm eV}$,
$\omega_{t}=2.297~{\rm eV}$) and DP2 ($\omega_{\tau}=2.429~{\rm eV}$,
$\omega_{t}=2.429~{\rm eV}$) reveal the middle polariton (MP) and the UP bands, respectively.
The low-intensity LP band cannot be reliably identified in
the SE 2DES spectrum, because 2DES peak intensities are proportional to the product of four relevant transition dipole moments, in contrast with  absorption spectra, which are proportional to the product of two transition dipole moments (see the discussion in Ref. \cite{Jindra15}).
The CP1 ($\omega_{\tau}=2.429~{\rm eV}$, $\omega_{t}=2.297~{\rm eV}$)
and CP2 ($\omega_{\tau}=2.297~{\rm eV}$, $\omega_{t}=2.429~{\rm eV}$)
quantify the coupling and polaritonic coherence between the states
MP and UP. A weak CP ($\omega_{\tau}=2.6~{\rm eV}$,
$\omega_{t}=2.429~{\rm eV}$) at $T_{w}=16$ fs reveals the phonon sideband.
The shape and intensity of the DPs and CPs depend significantly on
the waiting time $T_{w}$. The DPs at $T_{w}=0$ are predominately
elongated perpendicular to the main diagonal, but their elongation
turns into parallel at longer population times. A certain similarity
of the DPs at $T_{w}=0$ and 32 fs shows a significant contribution of the polaritonic beating with
a period of $\sim32~{\rm fs}$ revealing the energy difference of
$0.132\,{\rm eV}$ between the states MP and UP. The two negative
CPs are elongated parallel to the main diagonal. Similar to the DPs,
the CPs are also significantly modulated by the MP-UP energy-gap beating.
The overall decreases of the intensity of the SE contribution with
population time mirrors the CI-mediated population transfer from the
bright ${\rm S}_{1}^{(j)}$ states to the dark ${\rm TT}^{(j)}_1$ states and reflects
thereby the SF dynamics.

The GSB contribution is shown in panels (b) of Fig.~\ref{Fig4}.
It reveals the same pairs of the polaritonic DPs and CPs and, additionally,
a pair of weaker polariton-phonon CPs, ($\omega_{\tau}=2.6~{\rm eV}$,
$\omega_{t}=2.429~{\rm eV}$) and ($\omega_{\tau}=2.6~{\rm eV}$,
$\omega_{t}=2.297~{\rm eV}$). In comparison with the SE peaks, their
GSB counterparts look rather static. Nevertheless, the GSB contribution
reveals the vibrational wave-packet motion in the electronic ground
state, which is clearly manifested in the peak evolutions (see Ref. \cite{sun2}).

The ESA contributions are depicted in panels (c) of Fig.~\ref{Fig4}.
The SE and ESA contributions to 2DES spectra reveal photophysical
processes in the manifold $N=1$ \cite{skw20a}. However, the two contributions employ different reporter
states for delivering the same information: the SE employs the electronic
ground state $N=0$, while the ESA employs the doubly-excited manifold $N=2$.   Hence shapes and evolutions of predominantly positive SE contributions
and predominantly negative ESA contributions are quite similar. Nevertheless,
positions of the DPs and CPs in the SE and ESA spectra are slightly
shifted with respect to each other. At $T_{w}=0~{\rm fs}$, for example,
the (nearly) diagonal ESA peaks are located at ($\omega_{\tau}\approx2.297~{\rm eV}$,
$\omega_{t}\approx2.317~{\rm eV}$) and ($\omega_{\tau}\approx2.429~{\rm eV}$,
$\omega_{t}\approx2.416\,{\rm eV}$). All ESA peaks are elongated
parallel to the $\omega_{t}$ axis, due to a high density of states in the manifold $N=2$.   The peak ($\omega_{\tau}\approx2.297~{\rm eV}$,
$\omega_{t}\approx2.317~{\rm eV}$), which revels transitions from
the MP state to the states of the doubly-excited manifold, has the
highest intensity. Note also that the relative separation between
the DPs and CPs along the $\omega_{t}$ axis changes with $T_{w}$:
cf. the spectra at 0 fs with those at 16 fs. This
modulation is caused by the superposition of polaritonic and purely
vibrational tuning-mode oscillations. This reflects a higher complexity
of the ESA contribution, which involves polaritonic states with $N=1$ and 2.

Finally, the total 2DES signal $S(\omega_{\tau},T_{w},\omega_{t})$
is shown in Fig.~\ref{Fig4}(d). Since positions of the DPs and CPs
in the SE, GSB and ESA contributions nearly coincide, the total 2DES spectrum
exhibits positive, negative and dispersed peak shapes, the amplitudes
of which are determined by subtle cancellation effects in the superposition
of positive SE and GSB contributions and negative ESA contribution.
The overall structure of the total 2DES spectra is dominated by the
quartet of polaritonic DPs and CPs. In addition, a number of polariton-vibrational
peaks around $\omega_{\tau}\,\mathrm{or}\,\omega_{t}=$ 2.6 eV is seen at different $T_{w}.$ It is remarkable that
these hybrid polariton-vibrational peaks are more pronounced in the
total signal than in the SE, GSB, and ESA contributions --
owing to the aforementioned cancellation effects. The DPs are positive
while the CPs exhibit unusual asymmetry: the upper CP is always positive,
but the spectral structures around the lower CP can be negative (see the signals for $T_{w}=0$ and 16 fs).

\section{Simulation of polaritonic systems: unsolved problems, challenges, and outlooks}


\subsection {Model and Hamiltonian}

While the essential features of polariton dynamics can be effectively captured using the HTC Hamiltonian, the study of more practical SF-related photophysical processes has multifaceted significance and application value. The direct ${\rm S}^{(1)} \rightarrow {\rm TT}^{(1)}$  CI-driven SF is one of the possible mechanisms. In specific systems, charger transfer, locally-excited, doubly-excited, and multi-excitonic states facilitate the SF process, determine its pathways, and govern its efficiency \cite{MT18,MT19}. Furthermore, the coupling between these states can be of different nature, namely a dipolar coupling (which corresponds to the avoided crossing between the corresponding electronic/excitonic PESs) or a  vibrational-mode mediated coupling (which corresponds to the CI between the corresponding electronic/excitonic PESs). Obviously, embedding a multi-mode, multi-state system into a microcavity produces a cascade of coupled polaritonic states and PESs governing the SF process.  If several SF systems are embedded in one cavity, the situation becomes even more complex. It is essential though that all these polaritonic states directly or indirectly contribute to the coupling between the parent state ${\rm S}^{1}$ and the product state ${\rm TT}^{1}$. Hence, their net effect can be treated by introducing an effective ${\rm S}^{(1)}$-${\rm TT}^{(1)}$ coupling strength. We thus anticipate that the dynamical SF effects described here remain qualitatively valid if a large number of relevant electronic states is included.		
		
A proper characterization of cavity-mediated SF requires a careful selection of the relevant excitonic states and mechanisms of their couplings, notably for a large number $N_M$ of SF species embedded in the cavity. In this context, it is appropriate to mention  a recent discussion on the possibility to reach a strong-coupling regime in the cavities hosting a large number of species. Liu {\it et al.} pinpointed significant challenges in modifying photophysical dynamics in an optical microcavity, necessitating a molecular system with long excited-state lifetimes or architectures that enhance exciton-photon coupling strengths. These requirements are stipulated by the existence of the (quasi)degenerate manifold of dark ``reservoir" states (e.g., triplets) as well as the slow conversion from reservoir states to cavity polaritons~\cite{liu1}. In other words, the results of Liu {\it et al.} \cite{liu1}  as well as those of Takahashi and Watanabe \cite{Watanabe20} seem to demonstrate that a good portion of the singlet-state population bypasses the correlated triplet state and goes to the lower dark polaritonic states in the strong-coupling regime. On the other hand, a recent report by the Chergui group \cite{Chergui20} does not support this scenario. There appears to be a consensus that this challenge can be addressed through a careful and strategic redesign of the entire cavity QED system.
\cite{Menon24}.	

In excitonic systems, doubly-excited states are usually understood as the states in which two
chromophores are simultaneously excited. This interpretation, however, is not always sufficient for SF aggregates, because polyatomic molecules comprising these aggregates possess dense manifolds of higher-lying excited electronic states and the minimal system supporting the SF reaction is not a single chromophore, but a dimer. Usually, higher-lying electronic states do not contribute to the SF per se (see the discussion below though). Yet, optical transitions between the lower-lying and higher-lying  states affect dramatically  ESA contributions to spectroscopic signals. Hence  higher-lying states of individual molecules comprising SF dimers, aggregates, and materials have to be taken into account in any realistic simulation of spectroscopic responses.

Taking into account the  multimode (essentially, quasicontinuum) spectral densities of the microcavities \cite{Demler20,Gerrit22a,Ribeiro22} and finite cavity-mode lifetimes \cite{Huo24} may be essential for a proper description of the polaritonic dynamics in SF systems.
Thus, a single continuous-mode description of the system-cavity coupling may be insufficient or even inadequate.
However, the vertical excitation energy of a single dimer is around 2.2 eV. In this case, an effective single-mode description can be adopted for multi-mode cavities as only those modes in resonance with this transition would substantially contribute to the polaritonic SF. Similar effects have been explicitly demonstrated, e.g., in Ref.~\cite{GB24} where the fine (400 modes) description of the photonic spectral density was adopted.

It is crucial that all the additional states and specific inter-state couplings mentioned above can be effectively handled using the multi-D$_2$ Ansatz when required.	
Moreover, the methodology employing multi-D$_2$ Ansatz is efficient for arbitrary time-dependent Hamiltonians, non-Hermitian Hamiltonians, and models of multispecies, multimode bosonic baths \cite{zhao2,zhao1}.
Hence, the multi-D$_2$ Ansatz is a method of choice for handling quasicontinuous spectral densities of the cavity modes, finite lifetimes of the cavity/plasmonic modes, and fine  discretization schemes of the cavity modes.

Characterizing all relevant electronic states and parameterizing the SF Hamiltonians using ab initio quantum chemistry methods proves to be highly beneficial. \cite{MT18,MT19,YiZhao21,Haibo21,Vendrell23}.
In addition, {\it ab initio} approaches can also be applied to the entire systems of molecules/aggregates including the cavity. For example,
quantum mechanics/molecular mechanics (QM/MM) simulations of Refs.~\cite{Huo,Gerrit22,Gerrit22a} treat up to 512 polyatomic chromophores in a cavity.
Simulations such as those based on the {\it ab initio} QED \cite{Rubio23,Koch21}, the multiple spawning method \cite{Martinez24}, and the
cavity Born-Oppenheimer Hartree-Fock Ansatz \cite{Kowalewski23} can also provide valuable insights into dynamics of SF reactions in specific systems.

\subsection{Dynamics and spectroscopy}

Usually, accurate quantum simulations of up to two CI-carrying molecules/dimers in a cavity are performed \cite{Gu1,sun1,sun2}. {Vendrell \emph{et al.}, however, have simulated up to seven pyrazine molecules in a cavity, with each molecule having four modes and three states. They conducted a detailed analysis of the coupling between the cavity and the resonances of $S_0/S_2$ as well as $S_1/S_2$ \cite{Vendrell}.}
It is of great interest
    to push the simulations towards higher number of molecules/dimers $N_M$ and  explore how CIs affect cooperativity, e.g., to check the conventional $\sim \sqrt{N_M}$ scaling of the effective system-cavity coupling \cite{TAVIS,Huo22}.

To achieve this aim, a combination of advanced theoretical models and computational techniques appears promising. Few-molecule models with symmetry-optimized Hamiltonians \cite{JYZ23a} provide a simplified yet effective framework for understanding key interactions. Complementing these models, the multi-D$_2$ Ansatz, as utilized here \cite{zhao1,zhao2}, offers a powerful variational approach to accurately capture the dynamics of multi-species, multi-modes systems with a large number of DOFs. In particular, the HTC model containing dozens of qubits is readily tackled with the multi-D$_2$ Ansatz technique \cite{HTC}. Additionally, the multi-configuration time-dependent Hartree (MCTDH) method \cite{MT18,MT19,Vendrell19,Vendrell} stands out as a robust tool for exploring high-dimensional quantum dynamics, particularly in systems where vibrational and electronic degrees of freedom are strongly coupled. {The multi-layer extension of MCTDH has been carefully compared with the multi-$\rm D_2$ Ansatz method by Saalfrank and coworkers in a very recent study of vibrational relaxation dynamics at surfaces~\cite{Fischer}. While both rely on the time-dependent variational principle to arrive at an optimal wave function via solving highly nonlinear coupled equations of motion for variational parameters, the mDA method was found to be much more efficient (15,000s on a modern laptop) than the multilayer MCTDH (54,000s on a modern workstation). Furthermore, another potentially useful approach is the so-called time evolving density operator with orthogonal polynomials algorithm (TEDOPA)~\cite{Prior2010,Chin2011}. Through a basis change of the Hamiltonian, the TEDOPA method transforms the environment into a chain of harmonic oscillators with only local nearest-neighbor interactions\cite{Guimares,Tamascelli}. By utilizing Matrix Product States (MPS) for simulating the dynamics of one-dimensional quantum systems, TEDOPA effectively circumvents the otherwise unmanageable scaling associated with system size. Yet, the multi-$\rm D_2$ Ansatz method seems to be more flexible; for example, it has been applied very recently to the two-dimensional t-J model \cite{shen2024}. Furthermore, the multi-$\rm D_2$ Ansatz method can also address systems with long-range or strong interactions, as well as those involving entanglement. These challenges typically require a higher bond dimension in the MPS, leading to a long simulation time with TEDOPA.
It is worth mentioning that the tensor-network techniques \cite{Vidal18,Keeling22} have emerged as an effective approach for handling the computational challenges associated with large-scale quantum systems.}

Recently emerging emulations of CIs on quantum simulators \cite{CIsimulations23a,CIsimulations23b,Schoelkopf23} may also be extended in the future to cavity-CI systems. Unfortunately, the development of QM/MM approaches and numerically accurate fully quantum methods has largely progressed in isolation, with limited integration or cross-validation. This separation creates a significant gap in leveraging the strengths of both methodologies. Fully quantum methods provide high-precision insights into the dynamics of SF systems but are computationally intensive. In contrast, QM/MM methods offer a more computationally feasible framework by combining quantum-level detail with classical mechanics approximations, making them suitable for simulating larger systems or processes. Given these complementary strengths, it is highly desirable to bridge this gap by systematically validating QM/MM methods against fully quantum calculations. Such a validation process would not only enhance the reliability of QM/MM approaches but also enable their broader application in scenarios where fully quantum simulations are computationally prohibitive.

Disorder is of paramount importance to any realistic modeling of CI-driven cavity-mediated SF systems, as defects and disorder in semiconducting molecular materials can give rise to the formation of CIs \cite{Chuntonov22} and affect polaritonic transport substantially \cite{Schachenmayer22,HTC}.
Temperature adjustment is essential, e.g., for activating CI-mediated SI in rubrene \cite{Miyata} and for catalysing
reactions in polaritonic systems \cite{Campos}. The multi-D$_2$ Ansatz protocol based on the TFD \cite{BG21} was recently applied to explore dynamics of the pristine RD \cite{skw,Sun20}. Subsequently, Hou {\it et al.}~utilized this approach to obtain the finite-temperature dynamics and the absorption spectra of HTC model~\cite{HouE24}, as referred to in Sect.~\ref{HouE}. Looking forward, combining the multi-D$_2$ Ansatz with the TFD method, exploration of cavity-controlled SF processes at finite temperatures is anticipated to uncover new mechanisms, including the impact of vibrational polarization on the SF process induced by strong vibrational coupling, the influence of nonlinear coupling terms (such as the Kerr effect) within the cavity, and the role of multi-mode cavities. These factors are expected to provide deeper insights into the interplay between cavity dynamics and SF processes, potentially revealing novel pathways for enhancing efficiency and control.
Apart from multi-D$_2$ Ansatz,  MCTDH and variants of tensor-network methods are also well-suited for conducting fully quantum simulations of realistic polaritonic CI-driven SF systems, as they enable efficient evaluation of nonlinear spectroscopic signals
 \cite{Santoro23,GB_JCTC} and -- by using the TFD method \cite{BG21} and the recipe of Ref.~\cite{GB21} -- effects of finite temperatures and static disorder can also be taken care of in a (relatively) computationally inexpensive manner.

Yet a microscopic and fully quantum description of spectroscopic experiments has to await future developments: The bottleneck is accounting for the large number of  system and cavity DOFs and necessity to perform simulations on relatively long (picosecond to microsecond) timescales.
An extra challenge is to identify spectroscopic signatures of polaritonic effects in nonlinear spectroscopic signals of polaritonic CI-driven SF systems. Hence, accurate theoretical simulations and interpretations of the existing  \cite{Takahashi,Watanabe20,Chergui20} and forthcoming experiments are demanding but rewarding.

\subsection{Mechanisms of polaritonic CI-driven SF}
There exist several methods that can potentially enhance the SF yield. {Conventional practices are based on the adjustment of the cavity mode for achieving resonance and strong coupling of the LP state with the triplet state. Moreover, Vendrell \emph{et al.}~discovered the possibility of using the UP state as an entry door for the cavity-assisted SF pathways in a rigid system, which leads to a much higher yield \cite{Vendrell3}.}

On the other hand, the geometric-phase effects may also be harnessed for engineering SF in CI-driven polaritonic  systems. The controlling mechanisms may be implemented through collective topological polaritonic phases \cite{JYZ23} that are akin to those in topological metamaterials \cite{Alu23}.
It is also worthwhile to explore interfacing of  light-induced CIs in cavities \cite{Vibok22a,Vibok22b,Huo21} with intrinsic CI systems, as their combination may facilitate fine-tuning of the multiple CI landscapes \cite{VendrellVibok}.

Apart from the aforementioned cavity-related mechanisms, there exist other handles for tweaking SF, which are based on the optimization of the bias voltage \cite{Xiong18}, temperature \cite{Miyata,Xiong15,Wasielewski21}, carrier frequency \cite{BY22} and  intensity  \cite{Asahi16,Bardeen18a} of the
excitation pulse, or even static disorder~\cite{Chavez,Disorder24}.

Yet, it is not always clear whether
the coupling to the cavity (perhaps in combination with other methods mentioned above) enhances or suppresses the SF yield. It is  promising, therefore, to interface the existing  protocols for dynamic simulations of polaritonic CI-driven SF systems with quantum control routines \cite{Control1,Control2}, in order to find a robust pathway to comprehensive optimization of SF in cavities.

\section{Conclusions}


Integrating molecular materials into optical cavities has ushered in a new era of controlling quantum dynamics in complex systems. By coupling molecular excitations with quantized light fields, cavity environments reshape exciton energies, open new reaction pathways, and offer routes to surpass traditional photoconversion limits. This Perspective highlights recent developments in fully quantum simulations to understand CI-driven SF process in an optical cavity.
A numerically accurate Davydov trial state, the multi-D$_2$ Ansatz, is employed in a time-dependent variational framework
to explore dynamic and spectroscopic responses of the TC model, the HTC model, and
the CI-driven SF RD system strongly coupled to a cavity mode. For the disordered TC model, time evolution of the photon
population and the optical absorption spectra are also calculated by using the Green's function method to corroborate numerically accurate results of the multi-D$_2$ Ansatz. The impacts of temperature, static disorder, photon loss, and qubit-phonon coupling on the dynamics and absorption spectra of the HTC model are investigated by combining the multi-D$_2$ Ansatz combined with the TFD method.

Several microscopic SF mechanisms have been uncovered, viz.~proximity of the CI-driven polaritonic wave packet to the Franck-Condon region, cavity-induced  enchantment/weakening/suppression of SF, polaron/polariton decoupling, localization of the singlet-state population via engineering of the cavity-mode excitation, collective enhancement of SF.

Simulation and inspection of nonlinear femtosecond spectroscopic signals of the polaritonic CI-SF systems provides accurate information on real-time singlet-to-triplet transfer. In particular, 2DES spectra of the RD system are dominated by the quartet of DPs and CPs that reveal the MP and UP
bands, while the hybrid polariton-vibrational CPs contain information on the UP sideband. This clear separation of the polaritonic and vibrational
peaks is a noticeable feature of the spectra obtained. In addition, simulated 2DES spectra exhibit multiple peak-cancellation effects, and their population-time evolution is accompanied by polariton-vibrational beatings.

The findings emphasize the transformative role of polaritonic states in controlling SF dynamics, highlighting key factors such as cavity mode tuning, Rabi splitting, and electron-phonon interactions. Moreover, the study underscores the importance of integrating advanced simulation techniques, such as the multi-D$_2$ Ansatz, to capture the complex interplay of electronic, vibrational, and photonic states in many-body systems.

Strictly speaking, the level of description of the CI-driven cavity-enhanced SF systems adopted in this perspective and the ensuing effects outlined above pertain to so-called picocavities, in which the strong-coupling regime can be achieved for a few  embedded molecules \cite{Bhuyan}. To validate these predictions, it will be worthwhile to perform ultrafast spectroscopic (perhaps polarization-sensitive) experiments on a series of well characterized SF systems  (such as dimers, trimers, and tetramers) in solution and in picocavities. Furthermore, the effects akin to those demonstrated in this Perspective can likely be found in more  realistic CI-driven SF systems containing a large number of SF species strongly coupled to multi-mode cavities.

It is concluded that microcavities can open up new avenues for improving efficiency and stability of the CI-mediated SF, and ultrafast nonlinear spectroscopic signals -- if interpreted with adequate theoretical support --  can provide a real-time guidance on the SF reaction pathways. Yet, polaritonic SF studies have not reached the level of maturity, and many problems await efficient and practical solutions.
Looking ahead, the combination of fully quantum simulations, insightful modeling tools, and experimental innovations can guide us toward next-generation SF materials and cavity designs. By addressing these unsolved problems and challenges, researchers have the potential to leverage CIs and cavity architectures for surpassing conventional efficiency limits in solar energy harvesting and paving the way for groundbreaking innovations in next-generation photovoltaics and other optoelectronic applications.

{
}

\section*{Acknowledgments}
The authors gratefully acknowledge the support of the Singapore Ministry of Education Academic Research Fund (Grant No. RG87/20). K.~Sun would also like to thank the Natural Science Foundation of Zhejiang Province (Grant No LY18A040005) for partial support. M. F. G. acknowledges support
from the National Natural Science Foundation of China
(Grant No. 22373028).

\section*{Conflict of Interest}
The authors of no conflicts to declare.

\section*{Data Availability}
The data that support the findings of this study are available from the corresponding author upon reasonable request.

\appendix

\section{Green's function approach to the TC model }\label{Green0}

Based on Eqs.(4-7), one can investigate the energy spectral structure and the dynamics of TC model.
Without static disorder, $\omega_n=\omega_0$ and $g_1=g_2=\cdots=g_N=\omega_{\rm R}/\sqrt{N}$. Hence the propagators read
\begin{eqnarray}\label{Green4}
&&\langle 1_c, {\bf 0}_{\rm qu}|\hat{U}(t)|1_c, {\bf 0}_{\rm qu} \rangle \nonumber\\
&=& \frac{1}{2\pi i}\int_{-\infty}^{+\infty}\mathcal{\hat{G}}_{1_c,1_c}(\omega+i0^+)e^{-i\omega t}d\omega \nonumber\\
&=&\frac{(E_+-\omega_0)e^{-iE_+t}-(E_--\omega_0)e^{-iE_-t}}{E_+-E_-},
\end{eqnarray}
\begin{eqnarray}\label{Green4N}
&&\langle 0_c, {1}_{\rm qu}^n|\hat{U}(t)|1_c, {\bf 0}_{\rm qu} \rangle \nonumber\\
&=& \frac{1}{2\pi i}\int_{-\infty}^{+\infty}\mathcal{\hat{G}}_{1_{\rm qu}^n,1_c}(\omega+i0^+)e^{-i\omega t}d\omega \nonumber\\
&=&\frac{\omega_{\rm R}/\sqrt{N}e^{-iE_+t}-\omega_{\rm R}/\sqrt{N}e^{-iE_-t}}{E_+-E_-}
\end{eqnarray}
(the last lines in Eqs.~(\ref{Green4}) and (\ref{Green4N}) are obtained
by the residue theory). Here $z=\omega+i0^+$   and the  eigenenergies  of the bright states of the system, $E_{\pm}=1/2(\omega_c+\omega_0)\pm1/2\sqrt{(\omega_c-\omega_0)^2+4\omega_{\rm R}^2}$, are obtained from the poles of $\mathcal{\hat{G}}_{1_c,1_c}(\omega+i0^+)$ corresponding to the eigenmodes  strongly coupled to the
cavity mode. The bright-state energy gap
\begin{equation}\label{EG}
\epsilon =E_+-E_-=\sqrt{(\omega_c-\omega_0)^2+4\omega_{\rm R}^2}
\end{equation}
 is therefore formed, and the remaining $N-1$ degenerate dark states, which  are  decoupled from the cavity mode, are located inside the gap.

 Eq.~(\ref{Green4}) yields the following expression for the population of the photonic mode:
\begin{align}\label{Pop0}
P_{\rm ph}(t)=|\langle 1_c, {\bf 0}_{\rm qu}|\hat{U}(t)|1_c, {\bf 0}_{\rm qu} \rangle|^2.
\end{align}
Furthermore, the absorption spectrum can be written as
\begin{eqnarray}\label{spectr}
F(\omega)&=&\frac{1}{\pi}{\Re}\int_{0}^{\infty}\langle 1_c, {\bf 0}_{\rm qu}|\hat{U}(t)|1_c, {\bf 0}_{\rm qu} \rangle e^{i\omega t}dt \nonumber\\
&=&2{\Re}\Big\{\frac{E_+-\omega_0}{E_+-E_-}\delta(\omega-E_+)\Big\}+\nonumber\\
&&2{\Re}\Big\{\frac{E_--\omega_0}{E_--E_+}\delta(\omega-E_-)\Big\}
\end{eqnarray}
(for simplicity, all transition dipole moments are set to 1).
To account for  dissipation and dephasing caused by the coupling of the system to the environment, we add the non-Hermitian phenomenological lifetime terms $-i\kappa|1_c,{\bf 0}_{\rm qu}\rangle\langle 1_c,{\bf 0}_{\rm qu}|-i\Gamma_{n}|0_c,1^{n}_{\rm qu}\rangle\langle 0_c,1^{n}_{\rm qu}|$ to the Hamiltonian of Eq.~(\ref{TC}). When $\Gamma_{n}=\Gamma$, we can easily extend Eq.~(\ref{Green4})  by replacing $\omega_c$ ($\omega_0$) with $\omega_c-i\kappa$ ($\omega_0-i\Gamma$), respectively. In this case, the absorption spectrum assumes the form
\begin{eqnarray}\label{spectr1}
F(\omega)&=&\frac{2}{\pi}{\Re}\Big\{\frac{E_+-\omega_0}{E_+-E_-}\frac{\Im(E_+)}{[\omega-{\Re}(E_+)]^2+\Im(E_+)^2}\Big\}+\nonumber\\
&&\frac{2}{\pi}{\Re}\Big\{\frac{E_--\omega_0}{E_--E_+}\frac{\Im(E_-)}{[\omega-{\Re}(E_-)]^2+\Im(E_-)^2}\Big\}.\nonumber\\
\end{eqnarray}
Here finite  widths of the absorption peaks are determined by the imaginary parts of the eigenenergies. The above results are valid for any number of qubits.

Let us discuss now the influence of static disorder. namely, the usual diagonal disorder in qubit energies $\omega_n$.
It assume that $M[\omega_n^j]=\omega_0$, where random energies $\omega_n^j$ are uniformly distributed in $[\omega_0-W/2,\omega_0+W/2]$ and $W$ are the disorder strengths. Here $M[\cdot]$ denotes the ensemble average over random realizations of $\omega_n^j$. It is worth mentioning that the analytical results can be obtained if the Gaussian distribution disorder is adopted~\cite{Gera}.

Evidently, disorder removes degeneracy of the eigenenergies. In this case,  generalizations  of Eqs.~(\ref{Green4}) and (\ref{Green4N}) for $j$th disorder realization read
\begin{eqnarray}\label{Green5}
\langle 1_c, {\bf 0}_{\rm qu}|\hat{U}(t)|1_c, {\bf 0}_{\rm qu} \rangle_j=\sum_{m=1}^{N+1}\frac{\prod_{n=1}^{N}(E_m^j-\omega_n^j)}{\prod_{m^{\prime}\neq m}(E_m^j-E_{m^{\prime}}^j)}e^{-iE_m^jt},\nonumber\\
\end{eqnarray}
\begin{eqnarray}\label{Green5a}
\langle 0_c, {1}_{\rm qu}^n|\hat{U}(t)|1_c, {\bf 0}_{\rm qu} \rangle_j=\sum_{m=1}^{N+1}\frac{g_n\prod_{k\neq n}(E_m^j-\omega_k^j)}{\prod_{m^{\prime}\neq m}(E_m^j-E_{m^{\prime}}^j)}e^{-iE_m^jt}.\nonumber\\
\end{eqnarray}
Here $E_m^j$ are the eigenenergies  which can be obtained by diagonalization of the Hamiltonian (\ref{TC}). The photonic population reads then
\begin{align}\label{Pop}
P_{\rm ph}^{\rm av}(t)=M\big[|\langle 1_c, {\bf 0}_{\rm qu}|\hat{U}(t)|1_c, {\bf 0}_{\rm qu} \rangle_j|^2\big],
\end{align}
while the absorption spectrum  becomes
\begin{eqnarray}\label{spectr1a}
F^{\rm av}(\omega)&=&M\Big[2{\Re}\Big\{\sum_{m=1}^{N+1}\frac{\prod_{n=1}^{N}(E_m^j-\omega_n^j)}{\prod_{m^{\prime}\neq m}(E_m^j-E_{m^{\prime}}^j)}\delta(\omega-E_m^j)\Big\}\Big]\nonumber\\
\end{eqnarray}
(the superscript "av" stays for averaged). The finite lifetimes are accounted for  by replacing
$$\delta(\omega-E_m^j) \rightarrow \frac{1}{\pi}\frac{\Im(E_m^j)}{[\omega-{\Re}(E_m^j)]^2+\Im(E_m^j)^2}.$$

\section{Time evolution of the multi-D$_2$ Ansatz}\label{md2}

The time evolution of variational parameters $\mu_i$ (i.e., $A_{mn}(t)$ and $f_{mk}(t)$) is obtained by the time-dependent variational principle
\begin{eqnarray}
\label{Euler}
\frac{d}{dt}\frac{\partial L}{\partial \dot{\mu}_{i}^{\ast}}-\frac{\partial L}{\partial \mu_{i}^{\ast}} = 0,
\end{eqnarray}
where the Lagrangian $L$ reads
\begin{eqnarray}
\label{Lagrangian}
L&=&\frac{i}{2}\left[\langle {\rm D}_{2}^{M}(t)|\frac{\overrightarrow{\partial}}{\partial t}|{\rm D}_{2}^{M}(t)\rangle
-\langle {\rm D}_{2}^{M}(t)|\frac{\overleftarrow{\partial}}{\partial t}|{\rm D}_{2}^{M}(t)\rangle\right]\nonumber\\
&&-\langle{\rm D}_{2}^{M}(t)|\hat{H}_{\rm HTC}|{\rm D}_{2}^{M}(t)\rangle.
\end{eqnarray}
If $M$ is sufficiently large, the multi-D$_2$ Ansatz converges to the numerically ``accurate" solution to the Schr\"{o}dinger equation.
In terms of the multi-$\mathrm{D}_2$ parameters, the time evolution of the photon-mode population can be evaluated as follows
\begin{eqnarray}
\label{Pop1}
P_{\rm ph}(t)=\sum_{m,m^{\prime}=1}^M A_{m1_c}^*A_{m^{\prime}1_c}S_{mm^{\prime}},
\end{eqnarray}
where the Debye-Waller factor is given by
\begin{eqnarray}
S_{mm^{\prime}}=e^{\sum_{k}\left\{-\frac{1}{2}\left(\left|f_{mk}\right|^{2}+\left|f_{m^{\prime}k}\right|^{2}\right)+f_{mk}^{\ast}f_{m^{\prime}k}\right\}}.
\end{eqnarray}
Using  the definition of Eq.~(\ref{spectr}), we obtain  the following  multi-$\rm D_2$ expression for the absorption spectrum:
\begin{eqnarray}
\label{spectr2}
F(\omega)&=&\frac{1}{\pi}\Re\int_0^{\infty} dt \sum_{m,m^{\prime}=1}^M A_{m^{\prime}1_c}^{*}(0)A_{m1_c}(t) e^{-(\gamma^{\prime} - i\omega) t}\nonumber\\
&& \times e^{\sum_k\left\{-\frac{1}{2}(|f_{mk}(t)|^2+|f_{m^{\prime}k}(0)|^2)+f_{m^{\prime}k}^*(0)f_{mk}(t)\right\} }\nonumber\\
\end{eqnarray}
where $\gamma^{\prime}$ is the electronic dephasing rate.

Nonlinear spectroscopy is another important aspect in the dynamics investigation of various exciton-phonon-photon systems, as it provides valuable information on a variety of correlation functions.
The information provided by the linear absorption spectra and 2D photon echo (PE) spectra gives direct knowledge on exciton-phonon interactions and on the dephasing and relaxation processes that is elusive in the output from the traditional 1D spectroscopy.
In order to simulate the 2D PE spectra, we have to consider the interaction between the system and the classical light field. The corresponding Hamiltonian is given by $\hat{H}_{\rm L}=-\left(\mathbf{E}\left(\mathbf{r},t\right)\cdot\hat{\mu}_{+}+\mathbf{E}^{*}\left(\mathbf{r},t\right)\cdot\hat{\mu}_{-}\right)$,
with $\mathbf{E}(r,t)$  being the time-dependent electric field of the applied pulse sequence,
\begin{eqnarray}
\mathbf{E}\left(\mathbf{r},t\right)	&= &	 \mathbf{E}_{1}\left(\mathbf{r},t\right)+\mathbf{E}_{2}\left(\mathbf{r},t\right)+\mathbf{E}_{3}\left(\mathbf{r},t\right ), \nonumber\\
\mathbf{E}_{1}\left(\mathbf{r},t\right)	&= &	 \mathbf{e}_{1}E_{1}\left(t-\tau_{1}\right)e^{i\mathbf{k}_{1}\cdot\mathbf{r}-i\omega_{1}t+i\phi_{1}},\nonumber\\
\mathbf{E}_{2}\left(\mathbf{r},t\right)	&=&	 \mathbf{e}_{2}E_{2}\left(t-\tau_{2}\right)e^{i\mathbf{k}_{2}\cdot\mathbf{r}-i\omega_{2}t+i\phi_{2}},\nonumber\\
\mathbf{E}_{3}\left(\mathbf{r},t\right)	&= &	 \mathbf{e}_{3}E_{3}\left(t-\tau_{3}\right)e^{i\mathbf{k}_{3}\cdot\mathbf{r}-i\omega_{3}t+i\phi_{3}},
\end{eqnarray}
where $\mathbf{e}_{a}$, $k_{a}$, $\omega_{a}$, $E_{a}\left(t\right)$, and $\phi_{a}\left(a=1,2,3\right)$ denote the polarization, the wave vector, the frequency, the dimensionless envelope, and the initial phase, respectively.
It is common to define the pulse arrival times in $\hat{H}_{\rm L}$ as $\tau_{1}=-T_{w}-\tau,\thinspace\tau_{2}=-T_{w},\thinspace\tau_{3}=0$ where $\tau$
 (the so-called coherence time) is the delay time between the second and the first pulse, and $T_{w}$
 (the so-called population time) is the delay time between third and second pulse. In the short pulse limit, we have $E_{a}\left(t\right)=E_{0}\delta\left(t\right)$.
Before the optical excitation $(t\ll-T-\tau)$, the system is assumed to be in its global ground state $\left|g\right\rangle \left|0\right\rangle _{\rm ph} $.
Adopting this factorized initial condition allows one to neglect correlations between the primary system and its environment.
Without the contribution of excited-state absorption, the PE third polarization $P^{(3)}(t)$ can be decomposed into four nonlinear response functions $R_{1 - 4}$, which form the theoretical bases for the simulations of different 2D spectra.
The third-order response is directly given by the response functions in the impulsive limit, which can be expressed through the multi-$\rm D_2$ parameters as follows~\cite{sun2}
\begin{widetext}
\begin{eqnarray}\label{Rp3}
R_{1}(\tau,T_{w},t)&=&	 \sum_{i,j}^{M}\sum_{n,n_{1},n_{2},n_{3}}(\mathbf{e}_{4}^{*}\cdot\mathbf{\mu}_{n_{2}}^{*})(\mathbf{e}_{1}\cdot\mathbf{\mu}_{n_{3}})(\mathbf{e}_{2}^{*}\cdot\mathbf{\mu}_{n}^{*})(\mathbf{e}_{3}\cdot\mathbf{\mu}_{n_{1}})\nonumber\\
&&\times A_{jn_{1}n}^{*}(T_{w})A_{in_{2}n_{3}}(\tau+T_{w}+t)e^{\sum_{q}f_{jqn}^{*}(T_{w})f_{iqn_{3}}(\tau+T_{w}+t)e^{i\omega_{q}t}}\nonumber\\
&&\times e^{-\frac{1}{2}\sum_{q}(|f_{jqn}(T_{w})|^2+|f_{iqn_{3}}(\tau+T_{w}+t)|^2)},\nonumber\\
R_{2}(\tau,T_{w},t)&=&	 \sum_{i,j}^{M}\sum_{n,n_{1},n_{2},n_{3}}(\mathbf{e}_{4}^{*}\cdot\mathbf{\mu}_{n_{2}}^{*})(\mathbf{e}_{1}^{*}\cdot\mathbf{\mu}_{n}^{*})(\mathbf{e}_{2}\cdot\mathbf{\mu}_{n_{3}})(\mathbf{e}_{3}\cdot\mathbf{\mu}_{n_{1}})\nonumber\\
&&\times A_{jn_{1}n}^{*}(\tau+T_{w})A_{in_{2}n_{3}}(T_{w}+t)e^{\sum_{q}f_{jqn}^{*}(\tau+T_{w})f_{iqn_{3}}(T_{w}+t)e^{i\omega_{q}t}}\nonumber\\
&&\times e^{-\frac{1}{2}\sum_{q}(|f_{jqn}(\tau+T_{w})|^2+|f_{iqn_{3}}(T_{w}+t)|^2)},\nonumber\\
R_{3}(\tau,T_{w},t)&=&	 \sum_{i,j}^{M}\sum_{n,n_{1},n_{2},n_{3}}(\mathbf{e}_{4}^{*}\cdot\mathbf{\mu}_{n_{2}}^{*})(\mathbf{e}_{1}^{*}\cdot\mathbf{\mu}_{n}^{*})(\mathbf{e}_{2}\cdot\mathbf{\mu}_{n_{1}})(\mathbf{e}_{3}\cdot\mathbf{\mu}_{n_{3}})\nonumber\\
&&\times A_{jn_{1}n}^{*}(\tau)A_{in_{2}n_{3}}(t)e^{\sum_{q}f_{jqn}^{*}(\tau)f_{iqn_{3}}(t)e^{i\omega_{q}\left(T_{w}+t\right)}}\nonumber\\
&&\times e^{-\frac{1}{2}\sum_{q}(|f_{jqn}(\tau)|^2+|f_{iqn_{3}}(t)|^2)},\nonumber\\
R_{4}(\tau,T_{w},t)&=&	 \sum_{i,j}^{M}\sum_{n,n_{1},n_{2},n_{3}}(\mathbf{e}_{4}^{*}\cdot\mathbf{\mu}_{n}^{*})(\mathbf{e}_{1}^{*}\cdot\mathbf{\mu}_{n_{3}}^{*})(\mathbf{e}_{2}\cdot\mathbf{\mu}_{n_{2}})(\mathbf{e}_{3}\cdot\mathbf{\mu}_{n_{1}})\nonumber\\
&&\times A_{jn_{1}n}^{*}(-t)A_{in_{2}n_{3}}(\tau)e^{\sum_{q}f_{jqn}^{*}(-t)f_{iqn_{3}}(\tau)e^{-i\omega_{q}T_{w}}}\nonumber\\
&&\times e^{-\frac{1}{2}\sum_{q}(|f_{jqn}(-t)|^2+|f_{iqn_{3}}(\tau)|^2)}.
\end{eqnarray}
\end{widetext}
Here $\mathbf{e}_{4}$  refers to the polarization of the local oscillator field, and $\mathbf{\mu}_n$ are the transition dipole moment vectors.
$A_{jn_{1}n}^{*}(t)$ represents the probability amplitude at time $t$ for the exciton at the state  $\left|n_{1}\right\rangle$ with the initial state $\left|n\right\rangle$ and multiplicity $j$, and
$f_{jqn}(t)$ is the corresponding phonon displacement, also starting from
$\left|n\right\rangle{\exp}\left\{ \sum_{q}^{N_{q}}\left[f_{jq}(0)\hat{b}_{q}^{\dagger}-f_{jq}^{*}(0)\hat{b}_{q}\right]\right\} \left|\mathbf{0}\right\rangle_{\rm ph}$.

Similarly, the higher excited-state response functions are given by~\cite{sun2}
\begin{widetext}
\begin{eqnarray}\label{response functions1}
R_1^*(\tau,T_w,t)&=&\sum_{i,j}^{M}\sum_{\mbox{\tiny$\begin{array}{c}
nn_1n_2\\
n_3mm_1\end{array}$}}(\textbf{e}_4^*\cdot\mathbf{\mu}_{n_1m}^*)(\textbf{e}_1^*\cdot\mathbf{\mu}_{n}^*)(\textbf{e}_2\cdot\mathbf{\mu}_{n_3})(\textbf{e}_3\cdot\mathbf{\mu}_{n_2m_1})\nonumber\\
&&A^{j*}_{mm(n_1n)}(0)A^i_{mm_1(n_2n_3)}(t)e^{\sum_qf^{j*}_{mm(n_1n),q}(0)f^i_{mm_1(n_2n_3),q}(t)}\nonumber\\
&&e^{-\frac 1 2\sum_q(|f^{j*}_{mm(n_1n),q}(0)|^2+|f^i_{mm_1(n_2n_3),q}(t)|^2)},\nonumber\\
R_2^*(\tau,T_w,t)&=&\sum_{i,j}^{M}\sum_{\mbox{\tiny$\begin{array}{c}
nn_1n_2\\
n_3mm_1\end{array}$}}(\textbf{e}_4^*\cdot\mathbf{\mu}_{n_1m}^*)(\textbf{e}_1\cdot\mathbf{\mu}_{n_3})(\textbf{e}_2^*\cdot\mathbf{\mu}_{n}^*)(\textbf{e}_3\cdot\mathbf{\mu}_{n_2m_1})\nonumber\\
&&A^{j\prime*}_{mm(n_1n)}(0)A^{i\prime}_{mm_1(n_2n_3)}(t)e^{\sum_qf^{j\prime*}_{mm(n_1n),q}(0)f^{i\prime}_{mm_1(n_2n_3),q}(t)}\nonumber\\
&&e^{-\frac 1 2\sum_q(|f^{j\prime*}_{mm(n_1n),q}(0)|^2+|f^{i\prime}_{mm_1(n_2n_3),q}(t)|^2)}. \nonumber\\
\end{eqnarray}
\end{widetext}
The initial amplitudes of the higher excited-state are $A^{j*}_{mm(n_1n)}(0)=A^{*}_{jn_1n}(\tau+T_w+t)$, $A^i_{mm_1(n_2n_3)}(0)=A_{in_2n_3}(T_w)$,
$A^{j\prime*}_{mm(n_1n)}(0)=A^{*}_{jn_1n}(t+T_w)$, and $A^{i\prime}_{mm_1(n_2n_3)}(0)=A_{in_2n_3}(\tau+T_w)$, and the corresponding phonon displacements
are $f^{j*}_{mm(n_1n),q}(0)=f^*_{jn_1q}(\tau+T_w+t)$, $f^i_{mm_1(n_2n_3),q}(0)=f_{in_2q}(T_w)$, $f^{j\prime*}_{mm(n_1n),q}(0)=f^*_{jn_1q}(t+T_w)$, and $f^{i\prime}_{mm_1(n_2n_3),q}(0)=f_{in_2q}(\tau+T_w)$.

The contributions from stimulated emission, ground-state bleach, and excited-state
absorption, are expressed, respectively, as,
\begin{widetext}
\begin{eqnarray}
S_{\rm SE}(\omega_{\tau},T_{w},\omega_{t})&=&\Re\int_{0}^{\infty}\int_{0}^{\infty}dtd\tau [R_{2}(\tau,T_{w},t)e^{-i\omega_{\tau}\tau+i\omega_{t}t}+R_{1}(\tau,T_{w},t)e^{i\omega_{\tau}\tau+i\omega_{t}t}]\nonumber\\
S_{\rm GSB}(\omega_{\tau},T_{w},\omega_{t})&=&\Re\int_{0}^{\infty}\int_{0}^{\infty}dtd\tau [R_{3}(\tau,T_{w},t)e^{-i\omega_{\tau}\tau+i\omega_{t}t}+R_{4}(\tau,T_{w},t)e^{i\omega_{\tau}\tau+i\omega_{t}t}]\nonumber\\
S_{\rm ESA}(\omega_{\tau},T_{w},\omega_{t})&=&-\Re\int_{0}^{\infty}\int_{0}^{\infty}dtd\tau [R_{1}^*(\tau,T_{w},t)e^{-i\omega_{\tau}\tau+i\omega_{t}t}+R_{2}^*(\tau,T_{w},t)e^{i\omega_{\tau}\tau+i\omega_{t}t}]
\end{eqnarray}
\end{widetext}
The correlated 2D spectrum is given by the sum of the three contributions
\begin{eqnarray}
S(\omega_{\tau},T_{w},\omega_{t})=&&S_{\rm SE}(\omega_{\tau},T_{w},\omega_{t})
+S_{\rm GSB}(\omega_{\tau},T_{w},\omega_{t})\nonumber\\
&&+S_{\rm ESA}(\omega_{\tau},T_{w},\omega_{t}).
\end{eqnarray}

\section{A pedestrian introduction to TFD}\label{tfd}

Introduced in the 1970s to study quantum mechanics at finite temperatures, the TFD method has since found applications in various problems.
We combine the TFD method with the approach of multi-D$_2$ trial states and use it to study the finite-temperature dynamics of the HTC model.
In the TFD framework,
to solve the Schr\"odinger equation under a certain initial condition, we assume that the system is in thermal equilibrium initially at a given temperature, for which the initial state can be written as
\begin{equation}\label{6}
|\psi(0)\rangle=\rho_{\mathrm{ph}}^{1 / 2}|I\rangle|n\rangle|\tilde{n}\rangle=|\mathbf{0}(\beta)\rangle|n\rangle|\tilde{n}\rangle .
\end{equation}
Here $\rho_{\mathrm{ph}}$ is the density matrix of bath DOFs, which is usually represented by a set of resonators. $\tilde{H}$ is called the tildian and represents a fictitious Hamiltonian operator that can be derived from the original operator ${H}$ using a well defined mathematical procedure. $|{\tilde{n}}\rangle$ are arbitrary basis vectors of the fictitious space. $|\mathbf{0}(\beta)\rangle$ is the temperature-dependent vacuum state and $|n\rangle(|\tilde{n}\rangle)$ is the basis vector of the excitonic part of the Hamiltonian $H(\tilde{H})$. The dimension of the TFD wave function $|\psi(t)\rangle$ in the basis set representation is $\left(N_{\mathrm{ex}} \times N_{\mathrm{ph}}\right)^2$ , $N_{\mathrm{ex}}\left(N_{\mathrm{ph}}\right )$ denotes the number of electron (vibrational) basis functions, and the square is due to the doubling of the DOFs. This is exactly the same dimension as the density matrix of the physical system described, and therefore the TFD can be applied directly. We note that Eq.~\eqref{6} introduces a temperature effect using only the vibrational degrees of freedom, which suggests that the tilded part of the excitonic Hamiltonian $(|\tilde{n}\rangle)$ can fall. This statement was rigorously proved by Borrelli and Gelin \cite{BG21} who showed that  the TFD  Schr\"{o}dinger equation assumes the form
\begin{equation}\label{3}
i \frac{\partial}{\partial t}|\psi(t)\rangle=\bar{H}|\psi(t)\rangle .
\end{equation}
 The corresponding initial conditions can be written as
\begin{equation}
|\psi(0)\rangle=\rho_{\mathrm{ph}}^{1 / 2}|I\rangle|n\rangle=|\mathbf{0}(\beta)\rangle|n\rangle .
\end{equation}
The resulting expectation $\langle Q(t)\rangle$ is consistent with the expectation calculated by the Schr\"odinger equation with the Hamiltonian measure $\hat{H}$ after rewriting by TFD. The redefined Hamiltonian operator $\bar{H}$ is
\begin{equation}
\bar{H}=H-\tilde{H}_{\mathrm{ph}} .
\end{equation}
where ${H}$ is defined in Eq.~\eqref{HTC}, and $\tilde{H}_{\mathrm{ph}}$ is any operator acting in the tilded vibrational space. It is worth noting that the particular form of $\tilde{H}_{\mathrm{ph}}$ does not affect the expectation value $\langle Q(t)\rangle$. For simplicity, we have chosen
\begin{equation}
\tilde{H}_{\mathrm{ph}}=\sum_q \omega_q \tilde{b}_q^{\dagger} \tilde{b}_q .
\end{equation}
We obtain finally the new Schr\"{o}dinger equation by the Bogoliubov variation of Eq.~\eqref{3}
\begin{equation}\label{4}
i \frac{\partial}{\partial t}|\varphi(t)\rangle=\bar{H}_\theta|\varphi(t)\rangle, \quad|\varphi(0)\rangle=|\mathbf{0}\rangle|n\rangle,
\end{equation}
The transformed Hamiltonian ${H}_\theta$ takes the form
\begin{equation}
\begin{aligned}
\bar{H}_\theta= & e^{i G} \bar{H} e^{-i G} \\
= & \omega_c {a}^{\dagger} {a}+\sum_{n=1}^N\left[\omega_n {\sigma}_n^{+} {\sigma}_n^{-}+g_n\left({a}^{\dagger} {\sigma}_n^{-}+{a} {\sigma}_n^{+}\right)\right]\\
&+\sum_{k} \omega_k\left(b_k^{\dagger} b_k-\tilde{b}_k^{\dagger} \tilde{b}_k\right) \\
& -\frac{\lambda}{\sqrt{N}} \sum_{k, n} \cosh \theta_k \omega_k {\sigma}_n^{+} {\sigma}_n^{-}\left(e^{i k n} b_k+e^{-i k n} b_k^{+}\right) \\
& -\frac{\lambda}{\sqrt{N}} \sum_{k,n} \sin \theta_k w_k \hat{\sigma}_n^{+} \hat{\sigma}_n^{-}\left(e^{i k n} \tilde{b}_k^{\dagger}+e^{-i k n} \tilde{b}_k\right) .\\
&
\end{aligned}
\end{equation}
Here $G$ is the Bogoliubov unitary transformation operator defined by
\begin{equation}
\begin{aligned}
G=G^{\dagger}=-i \sum_k \theta_k\left(b_k \tilde{b}_k-b_k^{\dagger} \tilde{b}^{\dagger}_k\right) .
\end{aligned}
\end{equation}
with
\begin{equation}
\theta_k=\operatorname{arctanh}\left(e^{-\beta \omega_k / 2}\right) .
\end{equation}
The temperature effect is thus taken into account by coupling the physical system to the fictitious tilde system through the temperature-dependent mixing angle $\theta_k$.\\
The following relations have been used:
\begin{equation}
\begin{aligned}
& e^{i G} b_k e^{-i G} =b_k \cosh \left(\theta_k\right)+\tilde{b}_k^{\dagger} \sinh \left(\theta_k\right) \\
& e^{i G} \tilde{b}_k e^{-i G}=\tilde{b}_k \cosh \left(\theta_k\right)+b_k^{\dagger} \sinh \left(\theta_k\right) \\
& e^{i G}\left(b_k^{\dagger} b_k-\tilde{b}_k^{\dagger} \tilde{b}_k\right) e^{-i G}  =b_k^{\dagger} b_k-\tilde{b}_k^{\dagger} \tilde{b}_k .
\end{aligned}
\end{equation}
In numerical calculations, the high-frequency modes can be dropped from the tilded Hamiltonian quantities even at room temperature owing to $\theta_k \ll 1 , \sinh \left(\theta_k\right) \approx 0$, and $\cosh \left(\theta_k\right) \approx 1$. Therefore, there is an additional reduction in dimensionality compared to the standard Liouville's equation.

After the introduction of the tilded phonons, the multi-D$_2$ Ansatz with time-varying components, defined originally in Eq.~(\ref{D2}), can be recast as
\begin{equation}
\begin{aligned}
|\tilde{\mathrm{D}}_2^M(t)\rangle=& \sum_n|n\rangle \sum_{m=1}^M A_{m n}(t) e^{\left(\sum_k f_{m k}(t) {b}_k^{\dagger}\right)}|0\rangle_{\mathrm{ph}}\\
& \times e^{\left(\sum_k \tilde{f}_{m k}(t) \tilde{b}_k^{\dagger}\right)}|\tilde{0}\rangle_{\mathrm{ph}}.
\end{aligned}
\label{TD2}
\end{equation}
Here the non-normalized Bargmann coherent state is adopted. The advantage is that it does not need the additional
conjugate coherent state parameter space in the calculations, which doubles the computing efficiency in comparison to the case of the normalized coherent state. $|\tilde{\rm{D}}_2^M(t)\rangle$ in Eq.~(\ref{TD2}) can be used to solve Eq.~(\ref{4}).

According to the variational parameters obtained by the multi-$\mathrm{D}_2$ method, the time evolution of the population of the photon modes can be expressed in the following form:
\begin{equation}\label{5}
P_{\mathrm{ph}}(t)=\sum_{m, m^{\prime}=1}^M A_{m 1_c}^* A_{m^{\prime} 1_c} S_{m m^{\prime}} .
\end{equation}
where the Debye-Waller factor is
\begin{equation}
S_{m n^{\prime}}=\exp \left[\sum_k f_{m k}^*(t) f_{m^{\prime} k}(t)+\sum_k\tilde{f}_{m k}^*(t) \tilde{f}_{m^{\prime} k}(t)\right] .
\end{equation}
The multi-$\mathrm{D}_2$ expression for the absorption spectrum is
\begin{equation}
\begin{aligned}
F(\omega)= & \frac{1}{\pi} \Re \int_0^{\infty} d t \sum_{m, m^{\prime}=1}^M A_{m^{\prime} 1_c}^*(0) A_{m 1_c}(t) e^{-\left(\gamma^{\prime}-i \omega\right) t} \\
& \times e^{\left[{\sum_k f_{m^{\prime} k}^*(0) f_{m^{\prime} k}(t)+\sum_k \tilde{f}_{m^{\prime} k}^*(0) \tilde{f}_{m^{\prime} k}(t)}\right]} .
\end{aligned}
\end{equation}

\end{document}